\documentclass[useAMS,usenatbib]{mn2e}
\usepackage{epsfig,amsmath,natbib}

\newcommand{\emodel}{$\rm{e}^{-\tau_{\nu}}$\,}
\newcommand{\TLAE}{$T^{\rm{LAE}}_{\alpha}$\,}
\newcommand{\TQSO}{$T^{\rm{QSO}}_{\alpha}$\,}
\newcommand{\xhi}{$x_{\rm{HI}}$\,}
\newcommand{\nhi}{$N_{\rm{HI}}$\,}

\def\prosima{$\; \buildrel \propto \over \sim \;$}
\def\be{\begin{equation}} 
\def\ee{\end{equation}}

\def\msun{{M_\odot}}

\def\HI{\hbox{H~$\scriptstyle\rm I\ $}} 
\def\HII{\hbox{H~$\scriptstyle\rm II\ $}}

\def\gsim{\lower.5ex\hbox{\gtsima}} 
\def\lsim{\lower.5ex\hbox{\ltsima}} \def\gtsima{$\; \buildrel > \over 
\sim \;$} \def\ltsima{$\; \buildrel < \over \sim \;$} \def\prosima{$\; 
\buildrel \propto \over \sim \;$} \def\gsim{\lower.5ex\hbox{\gtsima}} 
\def\lsim{\lower.5ex\hbox{\ltsima}} 
\def\simgt{\lower.5ex\hbox{\gtsima}} 
\def\simlt{\lower.5ex\hbox{\ltsima}} 
\def\simpr{\lower.5ex\hbox{\prosima}} \def\la{\lsim} \def\ga{\gsim}

\def\gtsima{$\; \buildrel > \over \sim \;$} 
\def\ltsima{$\; \buildrel < \over \sim \;$} 
\def\gsim{\lower.5ex\hbox{\gtsima}} 
\def\lsim{\lower.5ex\hbox{\ltsima}} 
\def\simgt{\lower.5ex\hbox{\gtsima}} 
\def\simlt{\lower.5ex\hbox{\ltsima}} 
\def\simpr{\lower.5ex\hbox{\prosima}} 
\def\la{\lsim} 
\def\ga{\gsim}

\def\msun{\,{\rm M_\odot}}

\def\E3{{\cal E}_{\rm g}^{III}}

\def\nat{Nature}

\def\mnras{MNRAS}
\def\apj{ApJ}
\def\apjl{ApJL}
\def\apjs{ApJS}
\def\aap{A\&A}

\def\araa{ARA\&A}
\def\aj{AJ}

\def\apss{Ap\&SS}               

\def\pasj{Pub. Astron. Soc. Japan}

\title[Joint constraints on cosmic reionization]{Joint Ly$\alpha$ emitters - quasars reionization
constraints}         
\author[Baek, Ferrara \& Semelin]
{S. Baek$^{1}$, A. Ferrara$^{1}$ \& B. Semelin$^{2,3}$\\
$^1$ Scuola Normale Superiore, Piazza dei Cavalieri 7, 56126 Pisa, Italy\\
$^2$ LERMA, Observatoire de Paris, UPMC, CNRS, 61 Av. de l'Observatoire, 75014 Paris, France\\
$^3$ Universit\'e Pierre et Marie Curie, 4 place Jussieu, 75005 Paris, France}
\begin{document}
\maketitle

\begin{abstract}
We present a novel method to investigate cosmic reionization, using joint spectral information on high redshift 
Lyman Alpha Emitters (LAE) and quasars (QSOs). Although LAEs have been proposed as reionization probes, their use is 
hampered by the fact their Ly$\alpha$ line is damped not only by intergalactic \HI but also internally by dust. Our
method allows to overcome such degeneracy. First, we carefully calibrate a reionization simulation with QSO absorption line
experiments. Then we identify LAEs ($L_{\alpha} \ge 10^{42.2}\, \text{erg}$ and EW$> 20$\,\AA ) in two simulation boxes at $z=5.7$ and 
$z=6.6$ and we build synthetic images/spectra of a prototypical LAE. The surface brightness maps show the presence of 
a scattering halo extending up to 150 kpc from the galaxy center.  For each LAE we then select a small 
box of $(10h^{-1}\text{Mpc})^3$ around it and derive the optical depth $\tau$ along three viewing axes. At redshift $5.7$, we find that 
the Ly$\alpha$ transmissivity \TLAE $\approx 0.25$, almost independent of the halo mass. This constancy arises from the conspiracy 
of two effects: (i) the intrinsic Ly$\alpha$ line width and (ii) the infall peculiar velocity. At higher redshift, $z=6.6$, where 
$\langle x_{HI} \rangle =0.25$  the transmissivity is instead largely set by the local \HI abundance and \TLAE consequently 
increases with halo mass, $M_h$, from 0.15 to 0.3. Although outflows are present, they are efficiently pressure-confined by infall in 
a small region around the LAE; hence they only marginally affect transmissivity. Finally, we cast LOS originating from background QSOs 
passing through foreground LAEs at different impact  parameters, and compute the quasar transmissivity (\TQSO). At small impact 
parameters, $d < 1 $ cMpc, a positive correlation between \TQSO and $M_h$ is found at $z=5.7$, which tends to become less pronounced 
(i.e. flatter) at larger distances. Quantitatively, a roughly 10$\times$  increase (from $5\times 10^{-3}$ to $6\times 10^{-2}$) of \TQSO is observed 
in the range $\log M_h = (10.4-11.6)$. This correlation becomes even stronger at $z=6.6$. By cross-correlating \TLAE and \TQSO, we can 
obtain a \HI density estimate unaffected by dust. At $z= 5.7$, the cross-correlation is relatively weak,
whereas at $z = 6.6$ we find a clear positive correlation. We conclude by briefly discussing the perspectives for the application
of the method to existing and forthcoming data. 
\end{abstract}


\section{Introduction}
Lyman Alpha Emitters (LAEs) are galaxies showing prominent emission in the $2p \rightarrow 1s$ transition of the 
 hydrogen atom
resulting in a  Ly$\alpha$ emission line at $\lambda_\alpha = 1215.668$ \AA. At  redshift $z\gsim2$ the line can be observed in optical/IR
bands, thus allowing searches of distant galaxies traced by such radiation.  What is the source of  Ly$\alpha$  photons? Short-lived,
massive stars produce large amounts of UV photons with energy $h\nu >
13.6$ eV = 1 Ryd which ionize the surrounding gas. H-atoms recombine on a short time scale in
the dense interstellar medium and $\approx$ 2/3 of the recombination cascade ends up in Ly$\alpha$ photons, powering the observed
luminosity. The rapid production of dust associated with star formation process, and most
noticeably at high redshift by supernova explosions  \citep{Todi01},
might however lead to important attenuation of the Ly$\alpha$ emission
line. In spite of this drawback, Ly$\alpha$ is still considered the best available tracer of high-redshift star-forming galaxies. 
This was realized already more than four decades ago \citep{Part67}, but a full-scale application of the method had to await for 
technological progresses and became routinely used only towards the end of the last century \citep{Hu98}.

Since then,  several hundreds of LAEs over redshift $z\gsim6$ have been   
detected using narrowband imaging and spectroscopy (e.g., \citealt{Hu98,Rhoa03,Malh04,
Tani05,Kash06,Ouch08,Ouch09b,Ouch10,Cast10,Pent11}). Redshift $z= 6$ corresponds the end of 
the epoch of reionization, the last major cosmic phase transition in which the gas turned from a 
neutral to an ionized state. Albeit the general features of the reionization process are now understood
and constrained by available data \citep{Chou06, Bolt07, Ilie09c, Mitr11, Mitr11b}, many physical details as e.g., 
the initial mass function (IMF), escape fraction of Ly$\alpha$,  UV, and ionizing continuum photons,
intrinsic source spectral energy distribution, and interstellar (ISM) and intergalactic (IGM) medium inhomogeneities, remain 
only vaguely known.  There are strong hopes that LAEs can allow us to clarify these issues as several
authors (e.g., \citealt{Mira98,Sant04,Haim05,Dijk07,Daya08,Daya09}) have pointed out.

However, before we can fully exploit the LAE potential in this sense,
one has to deal with the fact that reionization itself affects the propagation of  
Ly$\alpha$ photons. In fact, in addition to the aforementioned
presence of dust, interstellar and intergalactic \HI atoms produce an effective opacity
caused by their large scattering cross-section to Ly$\alpha$ photons. 
As a result, inferring the intrinsic Ly$\alpha$ luminosity of a source
from the observed one it often quite difficult.  Further complications arise:
\citet{Daya11} have shown that a degeneracy between the IGM ionization state and the clumping of  ISM dust is imprinted in 
the LAE visibility. Stated differently, a wide range of IGM \HI
fractions, \xhi, can reproduce the observed Ly$\alpha$ luminosity function 
with different Ly$\alpha$ escape fractions, $f_{\alpha}$, from the galaxy, i.e.  increased transmission through a more ionized IGM  
is compensated by higher dust absorption inside the galaxy.  As a result, disentangling $f_{\alpha}$ from the transmissivity through 
the IGM is very challenging.

To attack this problem we propose here a novel strategy.  \citet{Adel03,Adel05} introduced a new method to measure IGM metallicity and 
neutral fraction around a target galaxy using background QSOs or galaxies. Such sources are separated from the target galaxy by an
angular distance small enough that their absorption spectra trace the IGM around the galaxy. 
They measured the spatial distribution of metals and HI using the absorption spectra of background QSOs/galaxies at
redshift $2\lsim z \lsim3$.  We exploit this successful method to investigate the transmissivity of LAEs through the IGM during reionization.
Several tens of QSOs are detected at redshift higher than $z\gsim5.7$ (e.g.\citealt{Fan01,Fan03,Fan06,Song04,Will07,Will10}) and
new QSOs are still being discovered \citep{Mort11} in the United 
Kingdom Infrared Telescope (UKIRT) Infrared Deep Sky Survey (UKIDSS).\footnote{http://www.ukidss.org}
 With hundreds of LAEs in the samples at  $z\gsim5.7$, it is possible to 
derive joint constrains on the IGM during cosmic reionization.

The main goal of this paper is to break dust-HI the degeneracy affecting the visibility of LAEs by isolating through the additional
information provided by near-passing QSO lines of sight the IGM contribution to the damping of the line. To achieve
this goal we will rely on a set of high-resolution, radiative transfer cosmological simulations. 

In Section \ref{sec_NDF}, we describe how we fine tune the \HI density field before running Ly$\alpha$ radiative transfer.
First we obtain the \HI density field in a $(100 \,h^{-1} \text{Mpc})^3$ volume through UV radiative transfer simulations at 
redshift $z=5.7$.  We normalize the UV photoionization flux so that the  synthetic absorption spectra match  the 
observed Ly$\alpha$  transmissivity deduced from QSO targets. Next, in Section \ref{sec_RT}, we present our LAE model 
study, a numerical aspect of LAE for an individual halo. We present the transmissivities for both LAE and QSO and find a relation between them 
at different redshifts (Section \ref{sec_TRANS}). We conclude with a critical discussion of the results in Section \ref{sec_CON}.
In appendix, we compare two Ly$\alpha$ radiative transfer methods, i.e. a full radiative transfer and a simplified $e^{-\tau_{\nu}}$ model
often used in the literature. 

\section{Neutral hydrogen density field}
\label{sec_NDF}

The first step of the method consists in building an accurate description of the IGM \HI 
density field at the epoch at which high-redshift  LAE are observed. This is a crucial step as,
due to its resonant  radiation transfer of the Ly$\alpha$ line is very 
sensitive to the precise value of \xhi. The \HI distribution is obtained through cosmological simulations
described in the following; however, it is of outmost importance that the simulated results
are calibrated against available absorption line observations before they can be used for our purposes.
In the remainder of this Section, we describe how we take care of this aspect.

\begin{figure}
 \includegraphics[width=84mm]{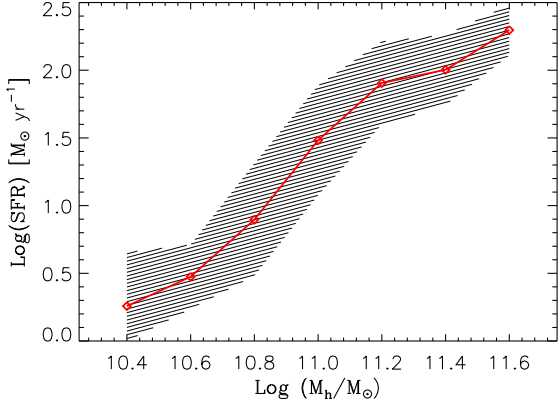}
 \caption{SFR as a function of the halo mass $M_h$. The red points are averaged over $M_h$, bins span
 0.2 dex, represented with the 1$\sigma$ error bar.} 
 \label{Mh_sfr}
\end{figure}

We start from two (modified, for reasons explained below) snapshots
(at $z=5.7, 6.6$) of the S1 cosmological reionization simulation
presented in \citet{Baek10} from which we take all relevant gas physical properties.
Such simulation results have been obtained by post-processing the
output of a GADGET-2 \citep{Spri05} cosmological hydrodynamical SPH 
simulation with the radiative transfer code LICORICE on an adaptive grid. 
Such grid has been constructed in such a way that each cell contains
at most $N=30$ SPH particles. This $N$ value  gives a minimum cell
size of 66.88 comoving kpc at z$\approx$6, which is smaller than the
size of a typical, virialized halo at z$\approx$6 ($\approx 200$ ckpc).  
The S1 run simulates a $(100h^{-1}\text{Mpc})^3$ volume with
$2\times256^3$ baryon and dark matter particles,  which gives a mass resolution of $3.2 \times 10^{9}\msun$ for dark matter and $6.9 \times 10^{8}\msun$  for baryons.
 We assumes WMAP3  cosmological parameters: $\Omega_m=0.24$,
$\Omega_b=0.042$, $h=0.73, \sigma_8=0.76$ \citep{Sper07}. 
 Reionization starts at $z\approx 14$ and ends at $z\approx 6$ as shown in Fig.2 in \cite{Baek10}.
Dark matter halos are identified with a friends-of-friends (FOF) algorithm. 
For each halo, we obtain its dark matter, $M_h$, and gas, $M_g$, mass and star 
formation rate (SFR) $\dot{M_*}$, with the method described in \citealt{Baek10}. 
 All details concerning the source modeling (SFR, IMF, SED, photoionization rate) are described in \citealt{Baek10}. We keep track
of stars formed inside each SPH particle according to a Schmidt-law,  so to at least qualitatively  account for photons produced by sub-resolution objects.
 If the star fraction of a particle is $>$0.1\% (corresponding to
  a mass of $\approx 10^{6}\,\msun$), we considered it as a UV source particle. The number
of source particles outside resolved halos is about 70\% in number,
corresponding to 20\% of the total luminosity of ionizing sources.
At $z=5.7$, the simulated mean star formation rate density is $\rho_{\star} = 8.8 \times 10^{-2}\,M_{\odot}\text{yr}^{-1}\text{Mpc}^{-3}$, 
in good agreement with current data \citep{Hopk06, Bouw11}.  
Fig. \ref{Mh_sfr} shows the SFR as a function of the halo mass $M_h$ of the galaxies identified as LAEs at  redshift $z=5.7$. 
A tight correlation between the two quantities is clearly observed, with a dependence of the SFR on halo mass approximately
as $\propto M_h^\beta$, with $\beta=1.9$ in the range $10.5 < M_h/M_\odot < 11.5$.
It is assumed that stars form according to a
Salpeter Initial Mass Function (IMF) in the range $1.6-120 M_\odot$.  The galaxy Spectral Energy Distribution (SED) has been calculated
by integrating the stellar properties given in \citet{Meyn05} and \citet{Hans94} over the IMF. 
We use 1000 frequency bins between 13.6 eV and 100 eV for stellar
type UV sources, and a photon packet propagates under periodic boundary condition.
We take into account supernova feedback by simply injecting in the surrounding gas particles an amount of $10^{48}$ erg per unit solar mass 
of stars formed of which 20\% (80\%) is in thermal (kinetic) form. All SPH neighbor particles are kicked with a velocity 
depending on the distance from the center. 
As we will explain later on, matching
the observed QSO transmissivity data, requires an ionizing photon escape fraction  $f_{esc}=0.09$.

\subsection{Calibration}

Even a small IGM \HI neutral fraction, $x_{HI}\approx  10^{-5}$ is sufficient to produce a Ly$\alpha$ scattering optical depth 
\be
\tau(z)=1.8\times 10^5 (\Omega_{M}h^2)^{-1/2}\left(\frac{\Omega_bh^2}{0.02}\right) \left(\frac{1+z}{7}\right)^{3/2} x_{HI}
\ee
larger than unity for mean IGM conditions at the redshift of main interest in this paper, $z = 5.7$.  Indeed the simulated volume-weighted
\HI neutral fraction is found to be in the range $-5 < \log \langle x_{HI} \rangle < -4$. Hence, due to the sensitivity of $\tau$
on \xhi it is necessary to determine the latter quantity with a precision $<10^{-5}$. 
The LICORICE code that we use for photo-ionizing radiative transfer uses a ray-tracing Monte Carlo method  on adaptive grid. 
The snapshot extracted from S1 shows Monte Carlo noise fluctuations on \xhi 
 which exceed the required precision. 
To overcome this problem and also to normalize the UV flux with QSOs observations by varying  $f_{esc}$, we post-process again the S1 snapshot. 
With an increased (by a factor $3300$) number of photon packets, corresponding to $10^{10}$ photon packets during a 10 Myr evolution
time; on average $\approx 10^6$ photon packets pass through each radiative transfer cell.  This procedure enables
us to achieve a precision on \xhi and hence to meet the required standard as shown in the next paragraph. 

\begin{figure}
 \includegraphics[width=84mm]{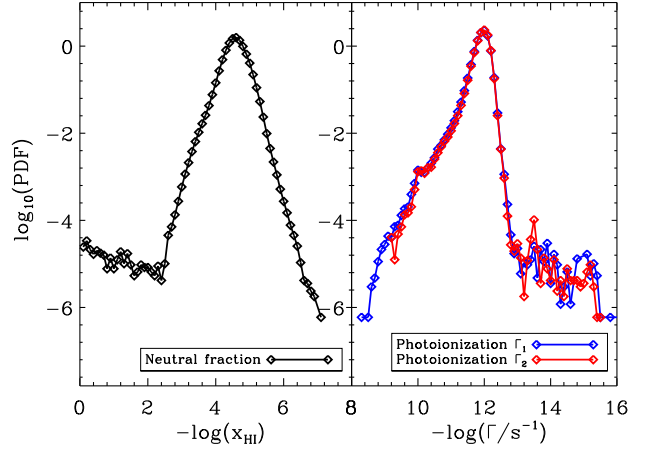}
 \caption{{\it Left panel:} Probability distribution function for the neutral fraction, \xhi. {\it Right:}  Same for the photoionization rate, $\Gamma$, 
computed either directly via full radiative transfer ($\Gamma _1$ curve) or by assuming photoionization equilibrium for each
particle  ($\Gamma _2$). The value of the ionizing photons escape fraction is $f_{esc}=0.09$.} 
 \label{NDF}
\end{figure}

\begin{figure}
\begin{center}
 \includegraphics[width=84mm]{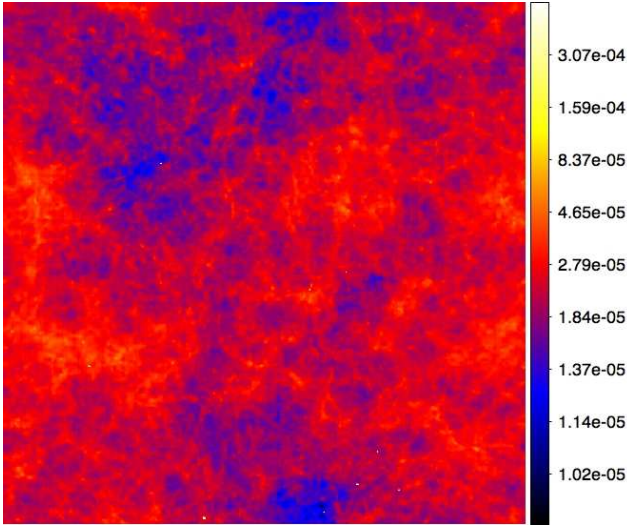}
 \caption{IGM Neutral fraction at $z=5.7$. The map is 100$h^{-1}$Mpc on a side and has
 a projected thickness of 33.33$h^{-1}$Mpc. The volume-weighted
 averaged neutral fraction is $\langle x_{HI}\rangle=2.3\times10^{-5}$.
The color scale is logarithmic.} 
 \label{nHI_field_119}
\end{center}
\end{figure}

\begin{figure}
 \includegraphics[width=84mm]{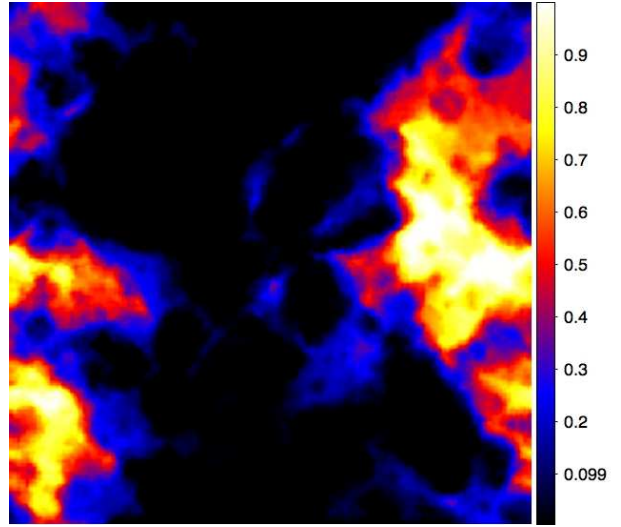}
 \caption{As in Fig. 4 for $z=6.6$. Here $\langle x_{HI}\rangle=0.25$. The color scale is logarithmic.} 
 \label{nHI_field_101}
\end{figure}

The two panels in Fig. \ref{NDF} show the probability distribution function (PDF) of  \xhi and  photoionization rate, $\Gamma$, in the high
precision run. The \xhi PDF has a peak at \xhi$\approx 3.2\times10^{-5}$
and closely approximates a log-normal distribution. The volume-weighted mean is $\langle$\xhi$\rangle=2.3\times10^{-5}$.
At this redshift, the reionization is already completed except for very high density clumps (visible as a high \xhi tail of the PDF) where the recombination rate is boosted. 
From the comparison of the $\Gamma$  PDFs in the right panel of Fig. \ref{NDF} we can also conclude that the IGM is in photoionization equilibrium. 
The blue curve, $\Gamma _{1}$, represents the photoionization rate distribution for the particles and it has been obtained from the full radiative transfer simulation. 
The red curve $\Gamma _{2}$, is instead the photoionization rate reconstructed from the simulated values of \xhi, gas temperature, $T$, and density, $n$, 
assuming ionization-recombination equilibrium, so that
\begin{equation}
\Gamma_2=\frac{(1-x_{\text{HI}})^2 n\alpha_{B}(T) }{x_{\text{HI}}},
\end{equation} 
where $\alpha_{B}(T)$ is the temperature-dependent recombination rate.
The volume-weighted photoionization rate is $\langle \Gamma \rangle =
1.14 \times 10^{-12} {\rm s}^{-1}$.
Both values $\langle x_{\text{HI}} \rangle$ and $\langle \Gamma \rangle$ are consistent
with the work of \citealt{Mitr11} which synthesizes and analyzes the observational constraints on reionization. 
Fig. \ref{nHI_field_119} and Fig. \ref{nHI_field_101} show the \xhi distributions  at $z =5.7, 6.6$, the latter obtained by rerunning
 the snapshot of S1 at $z=6.6$ with escape fraction $f_{esc}=0.09$ for comparison. At $z=5.7$, the simulation show that reionization is essentially 
complete and the neutral density field is relatively homogeneous. At
$z=6.6$, instead, large portions (25\%) of the simulated
volume are still essentially neutral 
and the ionization field is patchy with the HII regions confined around sites of intense galaxy/star formation (i.e., inside-out topology).

We caution that the mean \xhi we obtain at $z=6.6$ might be somewhat high. This is likely caused
by the fact that in the original S1 simulation, appreciable star formation appears only around $z=14$ (set by the 
simulation resolution). The resulting integrated e.s. optical depth is 0.062, to be compared with the WMAP7 fiducial
range observed value $0.088\pm 0.015$\citep{Jaro11}. On the other hand it is useful to remember that the Gunn-Peterson 
effect in high-$z$ quasars only gives lower limits for \xhi. The above tension does not represent a major issue for our 
study as in practice it only modifies slightly the relation between redshift and corresponding mean \xhi.


The volume-weighted mean $\langle x_{HI}\rangle=2.3\times10^{-5}$ we find is in excellent agreement with the data (e.g. \citealt{Fan06b}) ; however,
there is extra  information contained in the \xhi PDF distribution (Fig. \ref{NDF}, left panel) than can be used to calibrate the model. 
To this aim, we use the observed Ly$\alpha$ absorption line spectra of 17 $z\simgt 5.7$ QSOs from \citet{Song04} and \citet{Fan06b}. 
Let us define the transmission at a given redshift, $T(z_{\text{abs}})$ as the average ratio of observed flux $f_{\nu}$ to the intrinsic (unabsorbed) one:
\begin{equation}
T^{QSO}(z_{\text{abs}})\equiv \langle f^{obs}_{\nu}/f^{int}_{\nu} \rangle,
\label{eq_trans}
\end{equation}
where the average is over a certain wavelength interval along the line of sight (LOS).

In order to compute $T^{QSO}$ from our hydro simulations, we interpolate the \HI density field on a $512^3$ fixed grid, 
using the SPH smoothing length of each particle and compute the optical depth $\tau$. 
We produce 3000 synthetic spectra by piercing through the simulation box
with randomly oriented LOS, using the same spectral resolution ($\approx5300$) 
and wavelength interval (15\AA) to compute Eq. (\ref{eq_trans}) as assumed in the data (\citealt{Song04}).
The best match to the data is obtained by iteration, recursively varying the assumed $f_{esc}$ value. 

\begin{figure}
 \includegraphics[width=84mm]{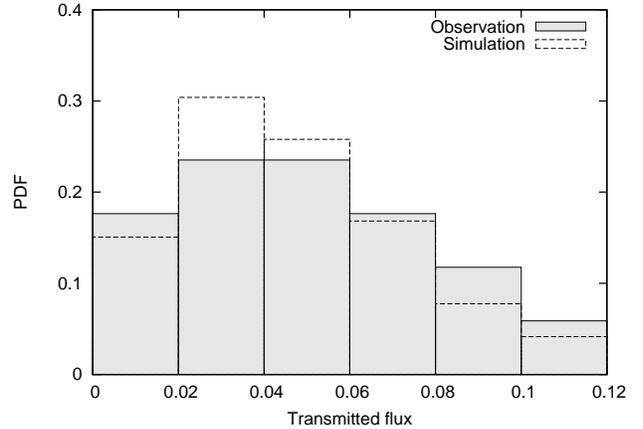}
 \caption{PDF of observed/simulated transmission of QSOs at $z=5.7$ with bin size 0.02. 
Filled box plot is from 17 observed QSOs in \citealt{Song04} and
 \citealt{Fan06b}. Dotted box plot is from 3000 synthetic spectra of QSOs.} 
 \label{histo_qso}
\end{figure}
In Fig. \ref{histo_qso}, we compare the LOS-averaged simulated $T^{QSO}$  with that obtained from 17 observed quasar 
spectra. Since the emitted photons are all on the blue side of the Ly$\alpha$ resonance, their transmissivity is very 
sensitive to the \HI distribution.  
The form of histogram varies sensitively with escape fraction; we found that the best fit with
$f_{esc}=0.09$ using our source model (initial mass function, spectral energy distribution, stellar
lifetime etc). We verified that for $ > 100$ synthetic spectra $T^{QSO}$ converges to the final distribution 
to $< 1$\%. This comparison assures that our modeling of the IGM \HI density field is accurate enough
to robustly predict the Ly$\alpha$ resonant line transfer from sources in the simulation volume.

\section{Simulating LAEs}
\label{sec_RT}
Ly$\alpha$ photons experience both frequency and spatial diffusion during the propagation and
modeling the Ly$\alpha$ spectrum emerging from galaxies requires a high level of accuracy
due to the resonant nature of the Ly$\alpha$ line. 
With increasing computational power, several groups have
developed Ly$\alpha$ transfer codes (e.g.\citealt{Zhen02,Ahn02,Dijk06,Verh06,Tasi06,Seme07,Pier09,Laur09,Fauc10}).
Most codes are based on Monte Carlo (MC) methods, capable of treating arbitrary geometries and various scattering
processes.
In this Section we describe our LAE model with radiative transfer and discuss numerical aspects
of the LAE modeling. Ly$\alpha$ radiative transfer in this work is done by LICORICE, which has been
widely tested and used in our previous works (\citealt{Seme07,Baek09,Baek10,Vonl11}).
However, differently from the ionizing UV radiative transfer, the Lyman line sector of LICORICE uses a fixed grid.

\subsection{LAE physical model}
We assume that the main source of Ly$\alpha$ emission in LAEs is star formation, and
compute the intrinsic luminosity of Ly$\alpha$ from their SFR (see Sec. \ref{sec_NDF}). 
About $2/3$ of ionizing photons are converted to Ly$\alpha$ photons assuming case-B recombination \citep{Oste89},
so the intrinsic Ly$\alpha$ luminosity can be expressed as,
\begin{equation}
L_{\alpha}^{int}=\frac{2}{3}(1-f_{esc})Qh\nu_{\alpha},
\end{equation}
where $Q$ is the ionizing photon production rate.
The intrinsic luminosity can be attenuated by dust absorption of Ly$\alpha$ photons in the LAE interstellar medium (ISM). 
As a result, only a fraction $f_{\alpha}L_{\alpha}^{int}$ can escape into the IGM.
Secondly, the frequency distribution of Ly$\alpha$ photons is broadened by Doppler effects induced by
galaxy rotation. As a consequence, the spectrum emerging from the LAE has the form
\begin{equation}
L_{\alpha}^{em}(\nu)=\frac{2}{3}Qh\nu_{\alpha}(1-f_{esc})f_{\alpha}\frac{1}{\sqrt{\pi}\Delta\nu_{D}}\text{exp}^{-(\nu-\nu_{\alpha})^2/\Delta\nu^2_{D}},
\label{Lya_eq}
\end{equation}
where $\Delta\nu_{D}=(v_c/c)\nu_{\alpha}$ and $v_c$ is the circular rotation velocity of the LAE.
If star formation is a quiescent process, it is $v_h\,\le \,v_c\,\le\,2v_h$ for realistic halo and disc properties (\citealt{Mo98,Sant04}), 
where $v_h$ is the halo rotational velocity at radius $r_{200}$ within which the collapsed region has a mean over density of 200 times the background.
We use $v_c=1.5v_h$ in this work. 
Finally, due to damping by \HI in the IGM, only a fraction, $T_{\alpha}^{LAE}$ of photons escaping out of the LAE
actually reaches us. Hence, the observed bolometric Ly$\alpha$ luminosity is,
\begin{equation}
L_{\alpha}^{obs}=L^{em}_{\alpha}T_{\alpha}^{LAE}.
\label{degeneracy}
\end{equation}
In the following we will fix $f_{\alpha}=1$, i.e. we neglect internal dust absorption.
This is because our aim is to isolate the effects of IGM on the Ly$\alpha$ transmissivity.
 A detailed study of
the effects of grains can be found in \citet{Daya11}. As already mentioned, they found a dust absorption/IGM
transmissivity  degeneracy that we aim at breaking with this work through an independent determination of 
$T_{\alpha}^{LAE}$. Before we proceed, we pause to discuss some interesting side issues.

\subsection{Spectral imaging}
\label{spec}

\begin{figure}
 \includegraphics[width=85mm]{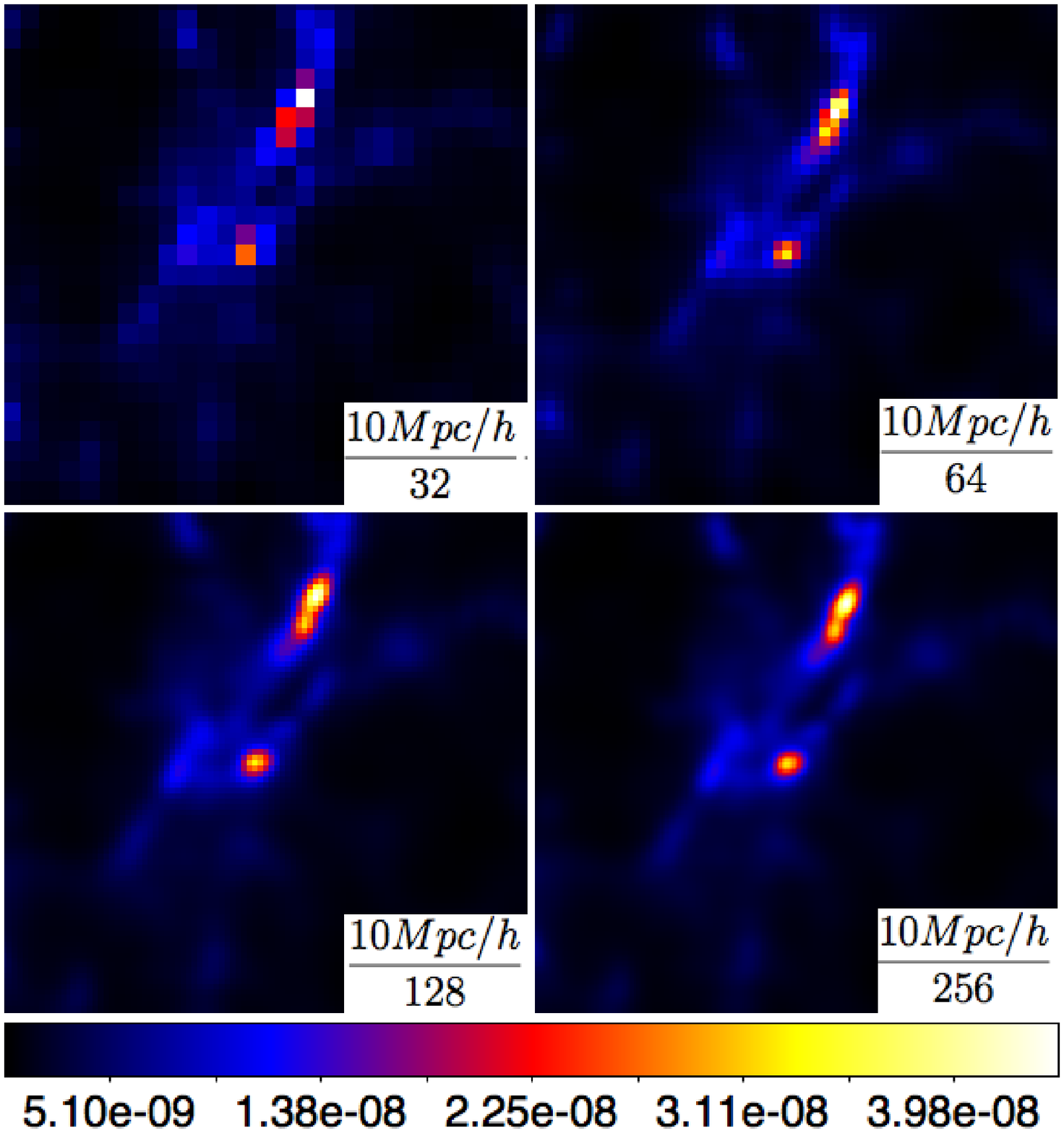}
  \includegraphics[width=85mm]{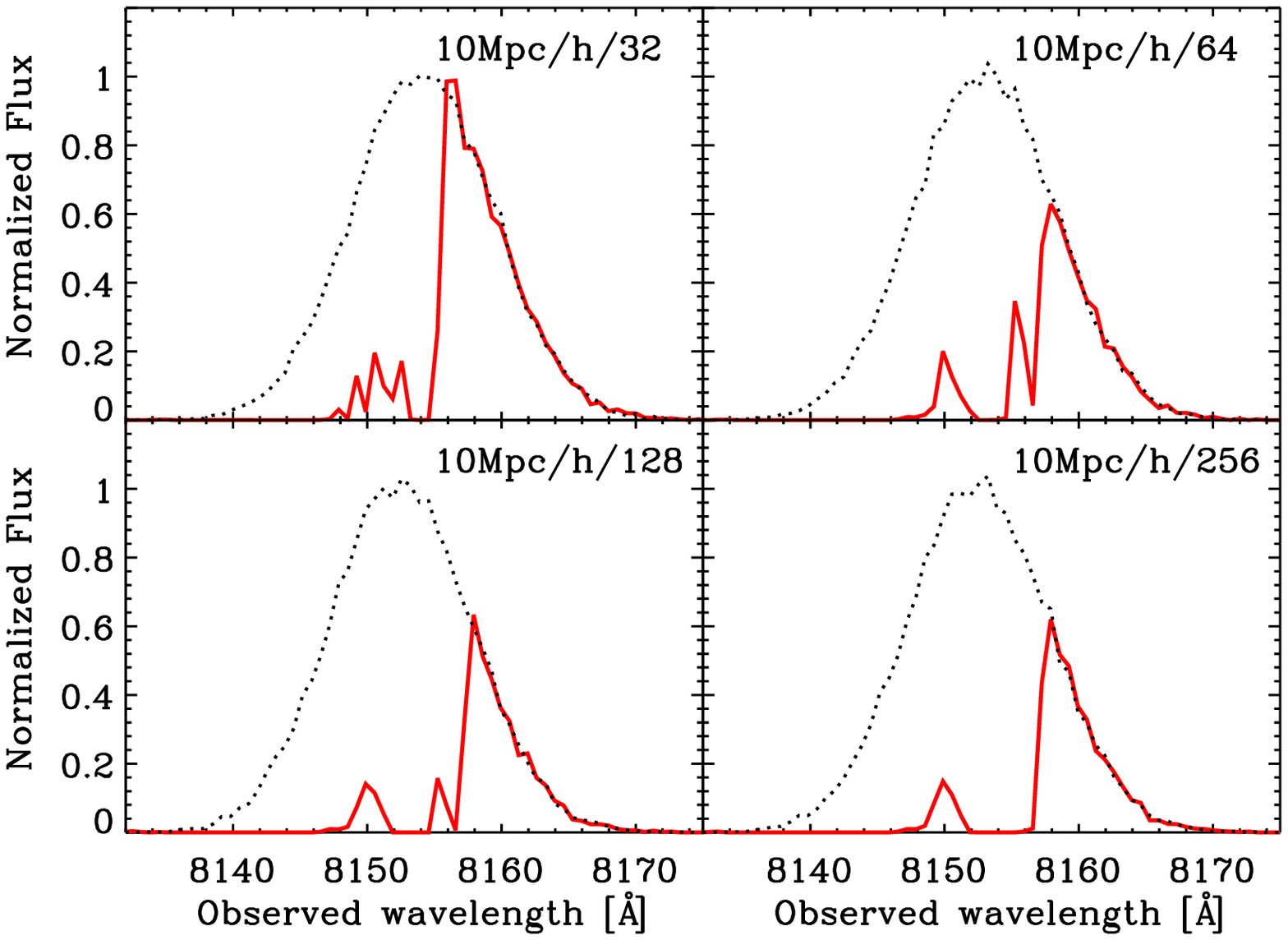}
 \caption{Neutral density field [$\text{cm}^{-3}$] at redshift $z=5.7$. The comoving pixel size is indicated in each panel. A
 LAE is at the center of the image.} 
 \label{resol}
\end{figure} 
In order to produce images and spectra of LAEs, we store the data as a three-dimensional array
similarly to \cite{Zhen02}. Two dimensions correspond to the sky plane (thus giving imaging
information), while the third one represents the frequency axis (spectral information) like in an IFU data cube. 
From the beginning and at each scattering, we compute the probability, $P$, for the photon to escape 
along the direction perpendicular to the sky plane; the photon is then added to the element of the array corresponding to the projected position 
and frequency. This procedure is necessary because only a very small fraction of the Monte Carlo photons escape exactly 
in the direction of the image (i.e. observer).
The direction into which the photon is scattered is deduced from a dipolar phase function $W(\theta) \propto 1+\text{cos}^2\theta$ , 
where $\theta$ is the angle between the incident and outgoing directions. The probability is added
to the array with weight $\text{e}^{-\tau_{\nu}}(1+\mu^2)d\Omega$, where $d\Omega$ is the 
solid angle subtended by the image pixel size, and $\mu$ is the cosine of the angle between the incident photon and the direction
perpendicular to the image plane. The optical depth $\tau_{\nu}$ is computed from the gas density
along the LOS to the image, with the frequency $\nu$ that the photon would have if it had been scattered in
that direction. The surface brightness of each pixel of the constructed image is
\begin{equation}
SB_{pix}=\frac{L_{\alpha}^{bol}}{d^2_L\Omega_{pix}N_{ph}}\sum_{i=1}^{N_{ph}}\sum_{j=1}^{N_{scat}}\frac{3}{16\pi}[1+(\bf{k\cdot\bf{k'}})^2]\text{e}^{-\tau_{\nu}},
\label{SB}
\end{equation}7
  where $L_{\alpha}^{bol}=\int_{-\infty}^{\infty} L_{\alpha}^{em}(\nu)
  d\nu$  is the bolometric
luminosity, $d_L$ the luminosity distance to the galaxy, 
$N_{ph}$ the number of photon packets, $\bf{k}$ the unit vector of the incident photon and $\bf{k'}$
is the one to the observer/image. The sum is performed over all photons and all of their scatterings.
LICORICE uses the comoving frequency and varying expansion factor $a(t)$ as a function
of the propagation time $t$ of the photon \citep{Baek09}. This method is more accurate than the usual approach which uses
the same expansion factor in the whole cosmological simulation box and
avoids first order errors in $(\delta a/a)$, which are not negligible when the propagation time $t$ reaches several tens of Myr.
To determine frequency diffusion after each scattering, we take into
account peculiar velocities;  the Hubble flow is also included as explained above.

Galaxies with $L_{\alpha}\ge 10^{42.2}\rm{erg\,s^{-1}}$ and observed equivalent width $\text{EW}\ge20$ \AA\, 
are identified as LAEs.
For each identified LAE in the simulation box, we select a smaller
volume of $(10h^{-1}\text{Mpc})^3$ centered on it.
Such volume size is chosen in such a way that all Ly$\alpha$ photons on the blue side of the line
are redshifted and hence not scattered anymore;  we have checked that on sub-volumes $>(10h^{-1}\text{Mpc})^3$ 
$T_{\alpha}^{LAE}$ is not seen to vary any longer. 
We then interpolate all physical properties of the gas particles in a
grid of $128^3$ weighting on the kernel, $w$, i.e.
a spherically symmetric spline function which depends on the smoothing length $h$ of SPH simulation  (see \citealt{Mona85}).
A resolved halo contains several tens of SPH particles within 10-100 kpc, so we can obtain higher resolution
by interpolating on a finer grid as shown in Fig. \ref{resol}. 
 The SPH density and velocity distribution can be sensibly
improved by interpolating on progressively finer grid using SPH smoothing length. 
We check the total transmitted flux varying grid size as
shown in Fig. \ref{resol}. Above grid linear size $10h^{-1}\rm{Mpc}/128$, \TLAE begins to converge. 
The pixel size is 16 kpc (physical) 
or $2.7''$ at $z=5.7$;  the probabilities are collected in a spectral array of 240 bins
spaced by 0.1 \AA \ in wavelength, thus  yielding a total spectral image depth of 24 \AA\, (rest frame).

\subsection{A prototypical LAE}
\label{proto}
\begin{figure*}
 \includegraphics[width=50mm]{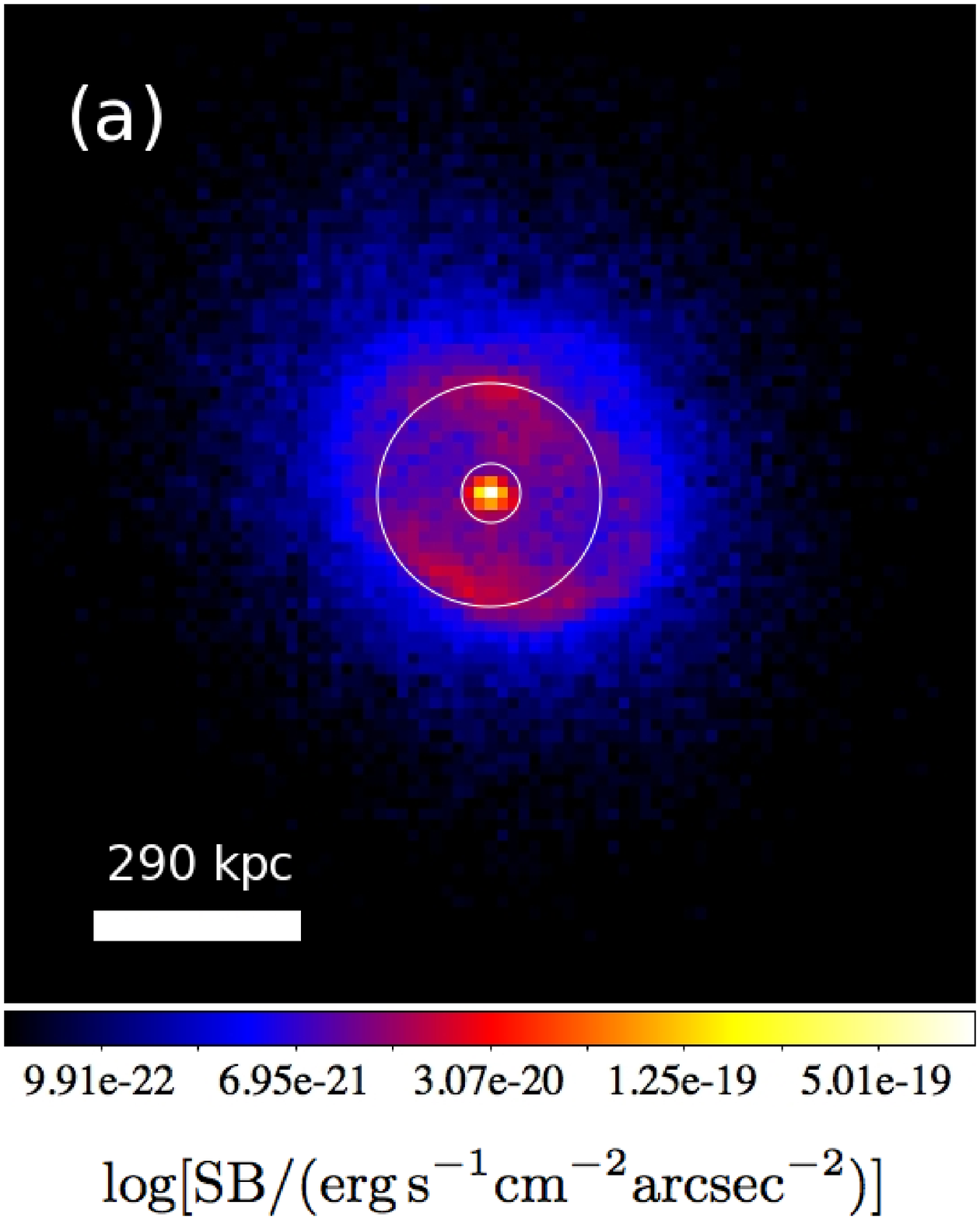}\,\,
  \includegraphics[width=50mm]{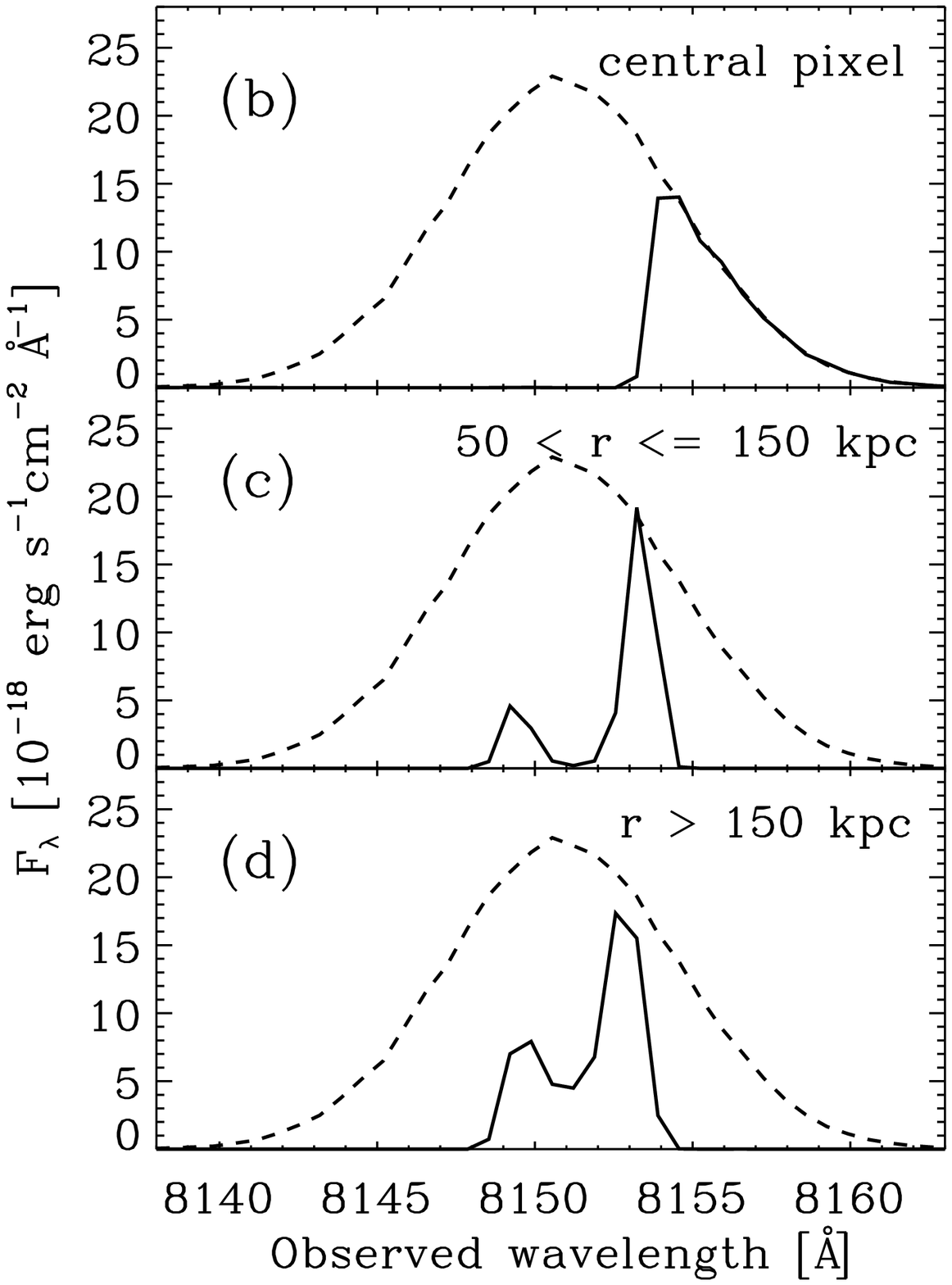}
 \includegraphics[width=60mm]{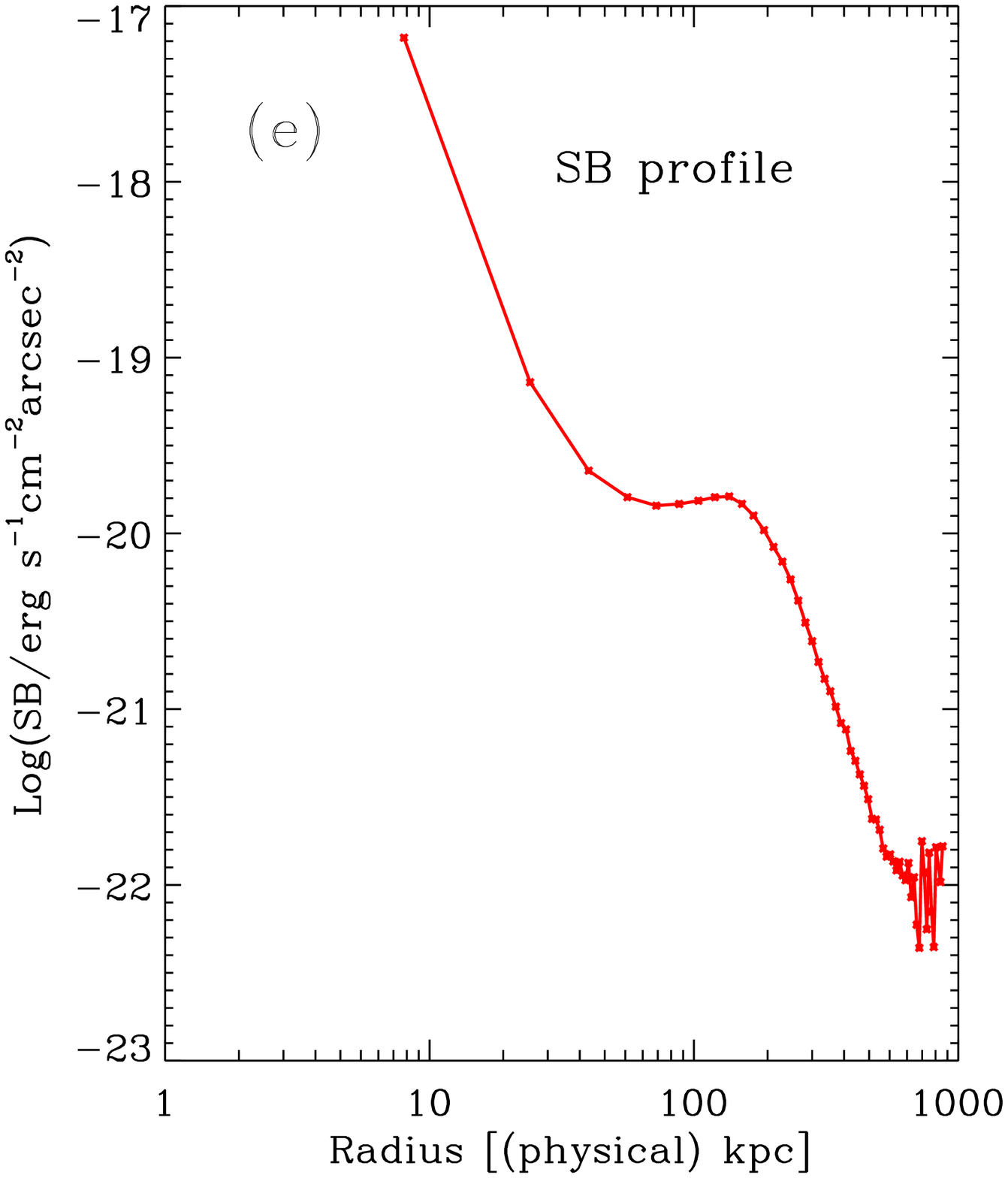}
 \caption{(a) Ly$\alpha$ surface brightness image of a prototypical LAE at $z= 5.7$. The two circles (of radius $r=50$ kpc and  $r=150$ kpc) 
denote the different regions for which integrated spectra are shown in the central panels.
Panels (b)-(d) show the  Ly$\alpha$ spectra, whose central frequency is redshifted at $\lambda_0=8150.55\, $\AA. The dotted curve in each panel shows the 
emerging Ly$\alpha$ profile, i.e. before any IGM damping. Solid curves refer to different regions:  (b) central pixel;  
 (c) $50 < r \leq 150$ kpc; (d) $r > 150$kpc. Panel (e) shows the spherically averaged Ly$\alpha$ surface brightness profile with bin size of 16 kpc.}
 \label{halo1739}
\end{figure*} 

To illustrate the physical features of a prototypical LAEat $z=5.7$  we select a halo with $M_h=8.6\times10^{10}M_{\odot}$, intrinsic luminosity 
$L^{int}_{\alpha}=7.3\times10^{43}\,\text{erg}\,\text{s}^{-1}$, and rotation velocity $v_c=200\,\text{km s}^{-1}$. These properties are consistent with 
those typically inferred for LAEs. The \HI column density integrated over the line-of-sight through the center of the simulation box ($10h^{-1}\text{Mpc}$) is 
\nhi$ \approx 6.5\times10^{16}\text{cm}^{-2}$. Assuming a static gas with homogeneous temperature $T=10^4$ K, 
this \nhi value gives a Ly$\alpha$ optical depth at line center $\tau_0\approx 1900$. 

Fig. \ref{halo1739} shows an overview of the results. Panel (a) is the Ly$\alpha$ surface brightness map obtained by collapsing 
data along the spectral array dimension. In practice, the current detection threshold of LAE narrowband imaging is 
$SB \approx 10^{-18}\text{erg}\,\text{s}^{-1}\text{cm}^{-2}\text{arcsec}^{-2}$ (e.g. \citealt{Shim06,Ouch08,Zhen10}). 
Only the central pixel ($SB=6.6\times10^{-18}\text{c.g.s.}$) exceeds such detection limit. This is because we are assuming a 
point source located at the galaxy center and therefore the photons reaching us are predominantly those escaping directly with little contribution from 
scattered radiation. At larger distances instead, all the flux is produced by Ly$\alpha$
scattering by \HI in the halo; its surface brightness dims with radius due to the
progressive decrease of \HI density.  The halo SB is potentially at reach of future instruments
and can be used to gain unique information about the IGM density structure around the emitter.

The spectra of the source at three different spatial locations are also shown in Fig. \ref{halo1739}.
The black dotted curve in each sub-panel shows the intrinsic Ly$\alpha$ profile emerging from the galaxy before 
being filtered by the IGM; the solid curves refer to the transmitted spectrum.  The spectrum of the central pixel
is very asymmetric as photons in the blue side of Ly$\alpha$ are completely suppressed as they are redshifted 
through the line core, while photons on the red side are fully transmitted beyond $\lambda > 8153$ \AA. 

Let us now analyze the spectra in two external regions, $50 < r \leq 150\,$kpc, and $r > 150$ kpc. The SB
observed at those locations is due to scattered photons escaping from the central pixel. Moving outward from the source, both  $n_{HI}$ 
and the magnitude of the peculiar velocity decrease with consequent reduction of the frequency diffusion; in addition,  the damping of 
the line core becomes less prominent. Both spectra show two sharp peaks in 
analogy with the case of an expanding sphere \citep{Seme07}; the left peak 
(blue side of Ly$\alpha$ line) is more suppressed than the right one because of Hubble expansion, which starts to dominate
over the infall velocity for distances $> 100$ kpc. The separation between the two peaks ($\approx 5 $\AA) is mainly set by the  gas velocity field 
rather than by the \HI density since \nhi is relatively small. In fact, with $T=10^4$ K and $\tau_0=1900$, the analytic solution  of \citet{Neuf90},
holding for a monochromatic source and an extremely optically thick system\footnote{See \citet{Dijk06} for the static, homogeneous sphere case.}, 
gives a separation of $\sim1$ \AA\, at redshift $z=5.7$, i.e. smaller than what is obtained here.

Panel (c) of Fig. \ref{halo1739} shows the spherically averaged Ly$\alpha$ SB profile as a function of galactocentric radius.  
The SB profile steeply ($SB \propto r^{-3}$) drops until $r\approx 50$ kpc, from where it flattens into a plateau 
extending to $r\approx 150$ kpc, followed by a final descent with slope similar to the inner one.  The SB dimming is
clearly related to the decrease of the \HI column density with radius. The presence of spherically symmetric  Ly$\alpha$ halos
around high redshift sources was pointed out by \citet{Loeb99}, who predicted the existence of a roughly uniform surface brightness
region of size $ \approx 0.1 r_{\star} =  0.67 ({\Omega_b}/{\Omega_M}) {\rm pMpc} = 100$ kpc for the cosmology adopted here.
Our simulation, allowing a more precise description of the \HI density distribution, is in rough agreement with the above conclusion if
the uniform surface brightness is identified with the plateau extension; the SB profile is however not constant within the Ly$\alpha$ 
halo as it steepens towards the center. 

The above agreement might be somewhat coincidental for the following reasons. In our simulation, the radius at which the Ly$\alpha$
optical depth approaches unity, hence allowing photons to free-stream away from the source, is $r_{\star}=500$ kpc, a factor 2 smaller
than predicted by the analytical model of \citet{Loeb99}, which was obtained by assuming a completely neutral, average density,  homogeneous 
gas with no peculiar velocity added to the Hubble flow. In contrast, our results show that  the typical \xhi in the halo is $10^{-5}-10^{-4}$ and the mean gas over density 
is $100-200$. If these values are used, $r_{\star}$ becomes much smaller than the analytical prediction. However, the further inclusion of
peculiar velocities, restores the rough agreement with the \citet{Loeb99} model. 
\begin{figure}
 \includegraphics[width=84mm]{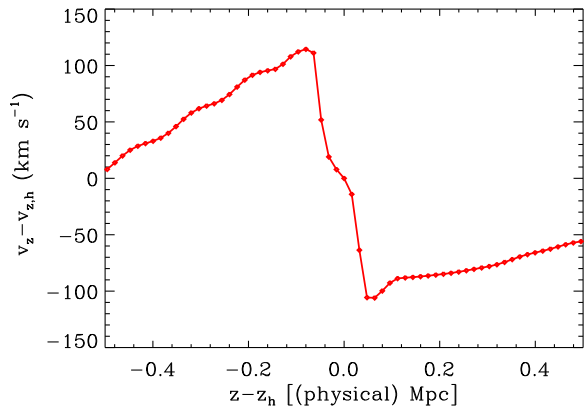}
 \caption{Peculier velocity (relative to the velocity of the central pixel) profile along the line passing through the central pixel along the
 line of sight (z-axis). Vertical dotted blue lines represent the limit of the Ly$\alpha$ diffusive area, from which photons enter free-streaming region.
   Velocity profile along z-axis is not symmetric because of other halos reside near the l.o.s.. 
   Positions of two halos (front/back with respect to the central halo) are indicated with blue circles, whose radius show relativistic
 mass size.} 
 \label{velo}
\end{figure} 

The previous argument demonstrates that infalling gas is the dominant process causing a reduction of transmissivity:
when infall is artificially turned off, in fact, transmissivity $T_{\alpha}^{LAE}$ for the selected LAE increases from 0.25 to 0.35. 
In principle, in addition to infall, additional peculiar velocities might be induced by outflows, which can modify the 
visibility of LAEs. For example, \citet{Dijk10} and \citet{Dijk11}, by using the spherically 
symmetric thin shell model introduced by \citet{Verh06,Verh08}, found that the transmissivity 
can be \gsim$5-10\%$ even through a fully neutral IGM. Although outflows are included in our simulations,
their effect on visibility  appears to be rather minor. The only visible effect of the galactic outflow emerging
from our prototypical LAE is to decelerate the infalling gas, as it can be seen from the change in the velocity
profile slope within the central 50 kpc (see Fig. \ref{velo}). Stated differently, it appears that the outflow 
expansion is severely quenched by the ram pressure of the infalling gas, thus affecting only mildly the final
transmissivity. A simple estimate reinforces this conclusion. The radius $R_e$ at which the infalling gas 
kinetic input rate becomes equal to the mechanical luminosity, $L_w$, of the outflow can be computed from the 
following equality
\begin{equation}
\frac{dM_{inf}}{dt} v_{inf}^2 = 4\pi R_e^2  \Delta \rho_{inf}(z)  v_{inf}^3 = \dot{M_*}\nu E_0 \epsilon_w = L_w,
\end{equation}
or
\begin{equation}
R_e = 70 \Delta ^{-1/2} \left(\frac{\dot{M_*}}{M_\odot {\rm yr}^{-1}}\right)^{1/2} \left(\frac{v_{inf}}{100 {\rm km s}^{-1}}\right)^{-3/2} {\rm kpc}. 
\end{equation}
In the previous equation $\Delta$ is the infalling gas over density with respect to the background density $\rho_{inf}(z) = \Delta \langle\rho_(z)\rangle$;
we have assumed 1 SN, of total energy $E_0=10^{51}$ erg, every 100 solar masses of stars formed and a kinetic energy conversion
efficiency $\epsilon_w = 0.1$. It is clear that the infall can quench the outflow on very small scales, and therefore its influence
on the escape of Ly$\alpha$ photons is marginal. Very likely though the
actual situation could be far more complex than it is possible to mode here: hydrodynamical instabilities might perturb the wind/infall
interface; the wind might be asymmetric and or bipolar. These complications require a dedicated investigation that is
beyond the scope of this study and we leave them for future work. 

The results shown in this Section are based on full radiative transfer of Ly$\alpha$ photons. However, this
procedure results in relatively large shot noise due to Monte Carlo sampling, making the comparison with
the QSO transmissivity (see below) challenging. For this reason in the following we will use  the $e^{-\tau_{\nu}}$
model, most often used in literature, to compute $T_{\alpha}^{\text{LAE}}$ rather than the full radiative transfer. We have performed a careful comparison
(discussed in the Appendix) between the two methods, from which we conclude that the differences
are not significant. Therefore we use the exponential model as the fiducial case in the rest of the paper.

 \section{Lyman Alpha Transmissivity}
\label{sec_TRANS}
The main aim of this paper is to put joint constraint on the LAE Ly$\alpha$ transmissivity using information
coming both from the LAE itself (i.e. in emission) and from the absorption spectra of background quasars. 
To perform this task, we cast thousands of LOS around each LAE halo to produce synthetic absorption QSO spectra 
with varying impact parameter. 
By analyzing these two sets of data in combination, we show in the following that it is possible to gain key information 
on the environment in which LAEs are embedded and to break the
dust absorption/HI scattering degeneracy that plagues the use of LAEs as reionization probes. We discuss the
properties of the Ly$\alpha$ transmissivity in LAEs and QSO separately first; next we combine the two result sets.

\subsection{Lyman Alpha Emitters}

Using the friends-of-friends (FOF) algorithm, we identified 3070 halos at redshift $z=5.7$, corresponding to
a halo number density $n=1.16\times 10^{-3}\,\text{Mpc}^{-3}$. For each halo, as explained above, we select a sub-volume
of $(10h^{-1}\text{Mpc})^3$  centered on it and interpolate all physical properties in a grid of $128^3$ cells. 
We use the \emodel model (see Appendix A)  and we average over 3 LOS along the $x,y,z$ axis from the halo center to the edge of simulation
box. We identified $\approx 2900$ objects meeting the LAE selection criterion  ($L^{int}_{\alpha}\ge10^{42.2}\text{erg}\,\text{s}^{-1}$ and EW$\ge20$\AA). 
\begin{table}
\centering
\small
\tabcolsep 3pt
\renewcommand\arraystretch{1.2}
   \begin{tabular}{c ||| c ||| c ||| c}
\hline
\hline
    Model   & RT method & PV &  Ly$\alpha$ profile    \\ 
\hline
\hline
    L1      & \emodel  & $\times$ & $v_c(M_h)$     \\
\hline
    L2      & full RT & $\surd$ & $v_c(M_h) $  \\
\hline
    L3     &  \emodel & $\surd$ & $v_c(M_h) $ \\
\hline
    L4    & \emodel & $\times $ & $v_c=\text{C}^{st}$   \\
\hline
    L5     & \emodel & $\surd $ &  $v_c(M_h) $ \\
\hline
\end{tabular}
\caption{Radiative transfer parameters. Model are represented by curves of different colors in Fig. \ref{LAE1}. PV is the
peculiar velocity of the gas. $v_c$ is the velocity of galactic rotation which determines Ly$\alpha$ line broadening.
(see Eq.\ref{Lya_eq})  }
\label{model}
\end{table}
\begin{figure}
 \includegraphics[width=84mm]{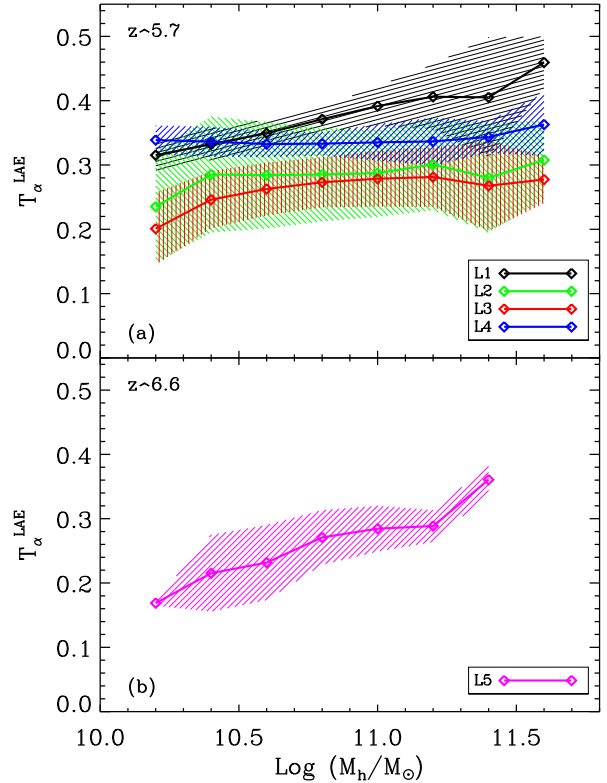}
 \caption{$T_{\alpha}^{\text{LAE}}$ as a function of  halo mass $M_h$ for identified LAEs. 
The halo mass bins span 0.2 dex and shaded area represent the 1$\sigma$ error bars in each mass bin.
Radiative transfer parameters for each model is described in Tab.\ref{model}. } 
 \label{LAE1}
\end{figure}
Fig. \ref{LAE1} shows $T_{\alpha}^{\text{LAE}}$
as a function of the identified LAE host halo mass $M_h$. The various radiative transfer parameters for each curve/model are
described in Tab. \ref{model}. \TLAE depends on several physical conditions, such as the LAE luminosity, 
clustering, Ly$\alpha$ profile width, gas infall, surrounding $n_{\text{HI}}$. The first three properties are positively correlated
with  \TLAE while the remaining ones tend to suppress \TLAE; the net effect therefore involves a complex interplay among them.
We run several simulations to understand how these elements affect \TLAE.
L3, implementing the \emodel model, is our fiducial run; L2 uses instead full radiative transfer. 
As discussed in Appendix A, transmissivities of L2 are slightly larger than L3 but show similar pattern. 

Except for the first three low-mass bins, \TLAE $= 0.2-0.3$, almost independent on halo mass.
This is an indication of a balance between transmissivity-enhancing and suppressing factors. For example, massive galaxies have larger SFR (see Fig. \ref{Mh_sfr}) 
and consequently higher photoionization rates. As a result, their \HII regions are larger and more transparent to Ly$\alpha$ photons (larger 
\TLAE). However, massive galaxies have also more prominent infall rates which tend to suppress transmissivity. This physical interpretation is supported by  
the results from run L1, in which we turned off peculiar velocity effects. Consistently with the above, \TLAE  increases with halo mass. 
However, we find that the positive correlation between \TLAE and $M_h$ is not as much driven by the increased photoionization rate, but rather by the broader 
intrinsic line widths. In fact, at $z= 5.7$  where the neutral HI density is low $\langle x_{\text{HI}} \rangle \approx 10^{-5}$ and very homogeneous, 
the ionizing flux from the LAE itself does not alter the ionization state of the IGM, and therefore the dependence on the halo mass is cancelled. Instead, if we   
run the L4 model without peculiar velocities {\it and} constant Ly$\alpha$ line width, we recover again a flat slope.
This is because broader Ly$\alpha$ line widths allow the escape of more red-wing photons. We conclude that \TLAE remains approximately 
constant with halo mass because peculiar velocities and  Ly$\alpha$ line width effects  roughly balance each other. 

For comparison, we run L5 at $z=6.6$ with same additional parameters as L3. At this redshift, reionization is still 
relatively patchy and the effects of the HII regions carved by the LAEs themselves becomes dominant. Hence, more massive galaxies are 
embedded in larger HII bubbles allowing a higher fraction of Ly$\alpha$ photons to reach us, i.e. \TLAE increases with halo mass.

\subsection{Quasars}

\begin{figure}
 \includegraphics[width=84mm]{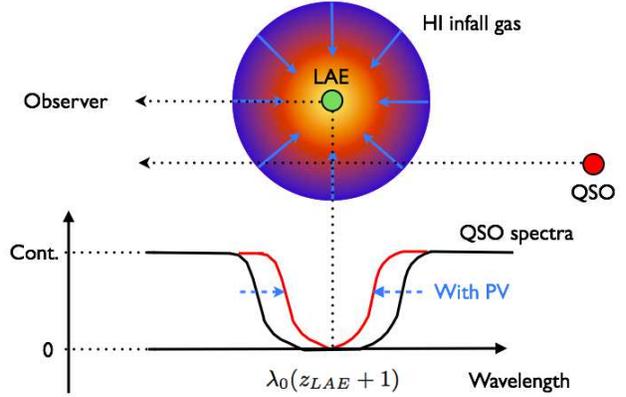}
 \caption{Schematic view of a QSO absorption spectrum whose LOS passes nearby an LAE. The red (black) spectrum is 
with (without) peculiar velocities. Due to infall, the core of the Voigt profile is shifted both to the  blue and red side 
of $\lambda_\alpha(1+z_{LAE})$ causing  \TQSO to increase.} 
 \label{infall_QSO}
\end{figure}

The next piece of information that we need to compute is the transmissivity, \TQSO, of QSOs whose LOS passes
through the environment of a foreground LAE at a given impact parameter, $d$ (see Fig. \ref{infall_QSO}).
\TQSO is computed following eq. \ref{eq_trans}, but we raised the spectral resolution to $R=40\,000$
to capture properties of the IGM around a LAE as much as possible.
At $R=40\,000$ one spectral bin size corresponds to $0.0304\,$\AA \, in the rest frame; for comparison,
at $z=5.7$, 1 comoving Mpc corresponds to $\Delta \lambda = 0.544 \, $\AA \,(rest frame)\, i.e. 17 spectral bins.
We select the transmitted flux of 17 spectral bins in the range $(\lambda_\alpha \pm \Delta \lambda/2)(1+z_{LAE})$
where $z_{LAE}$ is the redshift of the LAE. 
This range corresponds to positions in $ \pm 0.5$ cMpc around the LAE center along the LOS.
For each LAE, we cast 12,000 LOS along $(x,y,z)$ varying $d$, and compute averaged  values of \TQSO in the following 
four impact parameter ranges:  $d/{\rm cMpc} \in (0-0.5, 0.5-1, 1-2, 2-3)$.
\begin{figure}
 \includegraphics[width=84mm]{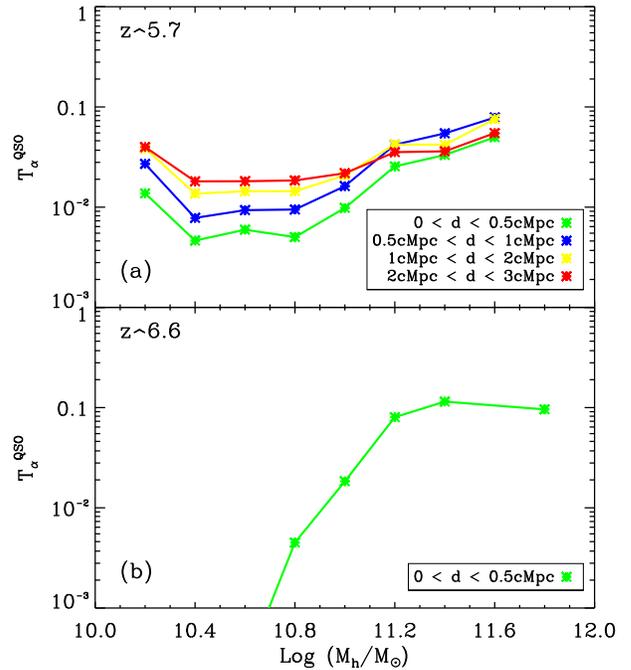}
 \caption{\TQSO as a function of halo mass $M_h$ for identified LAEs. The halo bins span 0.2 dex
 and curves represent median of each bin. } 
 \label{QSO0}
\end{figure}

Differently from \TLAE, at $z=5.7$ \TQSO shows a strong dependence on 
$M_h $ (Fig.\ref{QSO0}). This effect is caused by the gas infall around these halos, whose velocity 
increases rapidly with halo mass, and particularly beyond  $\log(M_h/M_{\odot})\approx10.8$. 
In addition, \TQSO for massive halos ($M_h\gsim 10^{11}M_{\odot}$)
is less dependent on impact parameter than for smaller halos.

At $z=6.6$, however, transmissivity drops sharply to $<10^{-3}$ below
$\log(M_h/M_{\odot})\approx11.2$, since reionization is not yet completed  
($\langle x_{\text{HI}} \rangle \approx 0.25$) and many smaller halos are still embedded in the dense neutral IGM.

As we have emphasized several times, gas peculiar velocities are an important physical factor in the determination 
of \TLAE. We define $v_{inf}$, the modulus of the infall velocity, as
\begin{equation}
v_{inf}=\frac{1}{N_{LOS}N_{PIX}}\sum_{i=1}^{N_{LOS}}\sum_{j=1}^{N_{PIX}}\sqrt{(v_j-v_{i,c})^2},
\end{equation}
where $N_{LOS}$ is the number of LOS, $N_{PIX}$ is the number of pixels around a LAE,  
$v_j$ is the peculiar velocity of the $j^{th}$ pixel  and $v_{i,c}$ is velocity of the central pixel which
contains the halo to the direction of the $i-th$ LOS.
We take $N_{LOS}=3$ along the three axis $(x,y,z)$  and $N_{PIX}=40$ which corresponds to 
$3h^{-1}$Mpc (comoving) around the halo, i.e. a few times the virial radius of typical LAEs, where 
gas infall should become evident. 
  \begin{figure}
 \includegraphics[width=84mm]{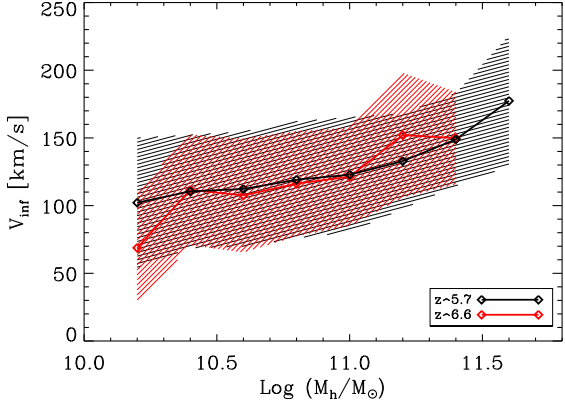}
 \caption{Gas infall velocity $v_{inf}$ in halos of mass $M_h$, represented with the 1$\sigma$ (68\%) distribution at each bin.} 
 \label{LAE3}
\end{figure}
Fig. \ref{LAE3} shows infall velocities of LAEs as a function of halo mass $M_h$ at redshift 5.7 and 6.6.
$v_{inf}$ increases with halo mass at both redshifts. Indeed, as we will see shortly, $v_{inf}$ is a key parameter to understand
\TQSO as well as \TLAE.

Photons from the QSO are in the blue side of Ly$\alpha$ line and they are completely scattered out of the
LOS by even tiny \HI amounts. Usually, the \nhi around a LAE is large enough to strongly suppress the quasar flux;
however,  peculiar velocities (infall and outflows) can shift the core of the Voigt profile and prevent the suppression.  
Different from \TLAE, a large infall velocity increases \TQSO since the core of the Voigt profile is shifted from the central Ly$\alpha$ frequency. 
As we show in Fig. \ref{infall_QSO}, photons in the red (blue) side of $\lambda_\alpha (1+z_{LAE})$
interact with the gas having a velocity component parallel (antiparallel) to their propagation direction. 
In both cases, the shift of the Voigt profile core produces a \TQSO increase. 

\begin{figure}
\includegraphics[width=84mm]{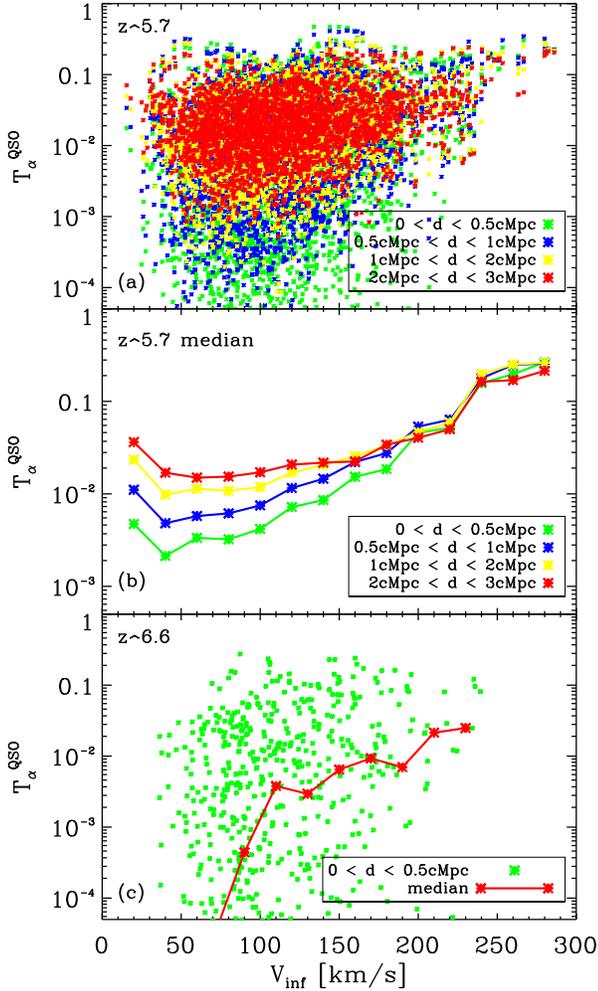}
\caption{\TQSO as a function of $v_{inf}$. (a) all data points, (b) the median of the distribution, (c) all data points and the median at redshift $z=6.6$. } 
\label{QSO1}
\end{figure}

Panel (a) of Fig. \ref{QSO1} shows  all data points for \TQSO 
at different impact parameters $d$, while panel (b) shows the median of the distribution, both as a function of $v_{inf}$.  
Values \TQSO$ < 5\times10^{-5}$ are not presented since they overshoot the observational precision ($10^{-4}$). 
In this case a  clear correlation between  \TQSO and $v_{inf}$ exists. 
The slope is sharpest at the smallest impact parameter and at large infall velocities.  
This is expected from the simple argument that the environmental properties of the gas deeper into the LAE
gravitational potential, where the component of the peculiar velocity along the LOS and \nhi are higher (see Fig.\ref{resol}), 
deviate more from the mean IGM ones. At larger impact parameters the slope of the median becomes flatter and closer 
to the mean IGM value $\approx 0.04$ that we obtained in the Sec. \ref{sec_NDF}. The medians become independent of $d$ 
beyond $v_{inf}\ge$200 $\rm{km\,s^{-1}}$, exceeding the mean IGM transmissivity by a factor of about 3.

For QSOs, the peculiar velocity is the most dominant factor to determine their transmissivity. \TQSO is computed 
in the wavelength range $\Delta \lambda=0.544$ \AA\,  (rest frame, covers the 1 cMpc bracketing the center)  
corresponding to a Doppler shift velocity of 135 $\rm{km\,s^{-1}}$. The magnitude of $v_{inf}$ for massive halo is 
about 150-200 $\rm{km\,s^{-1}}$, well above the velocity required to shift the core of the Voigt profile out of the above 
wavelength interval. Thus even if the \nhi is  highest within the smaller impact parameter ($0<d<0.5$ cMpc), the infall velocity 
is large enough to produce an increase of \TQSO.  Since the infall velocity is 
correlated with halo mass, we observe a similar tendency - the independence of \TQSO on impact  parameter - in panels  
(b) of Fig. \ref{QSO0} and Fig. \ref{QSO1}.

Panel (c) of Fig. \ref{QSO1} shows similar data for $z=6.6$ for the closest impact parameter, which shows a steeper positive correlation
between \TQSO and $v_{inf}$. 
At this redshift, 25\% of the volume is still neutral and some of the halos are embedded   
into completely neutral patches. As mentioned, at
higher redshift the global ionization state of the gas dominates the
transmissivity over peculiar velocity effects and the size of \HII
bubble plays a more important role on \TQSO than $v_{inf}$.
The data of \TQSO at $z=6.6$ is more scattered than the data at $z=5.7$ due to the inhomogeneous \nhi distribution 
and the smaller number of resolved halos. The median of \TQSO below $v_{inf}=120\rm{km\,s^{-1}}$ drops very sharply and even the 
median is under $10^{-4}$ for $v_{inf}\le75\rm{km\,s^{-1}}$.  We interpret this trend as due to the fact that halos with small $v_{inf}$ 
host less massive/luminous galaxies, possibly embedded in smaller HII regions leading to a complete \TQSO suppression.

\subsection{Lyman Alpha Emitters/Quasars correlation}

\begin{figure}
 \includegraphics[width=84mm]{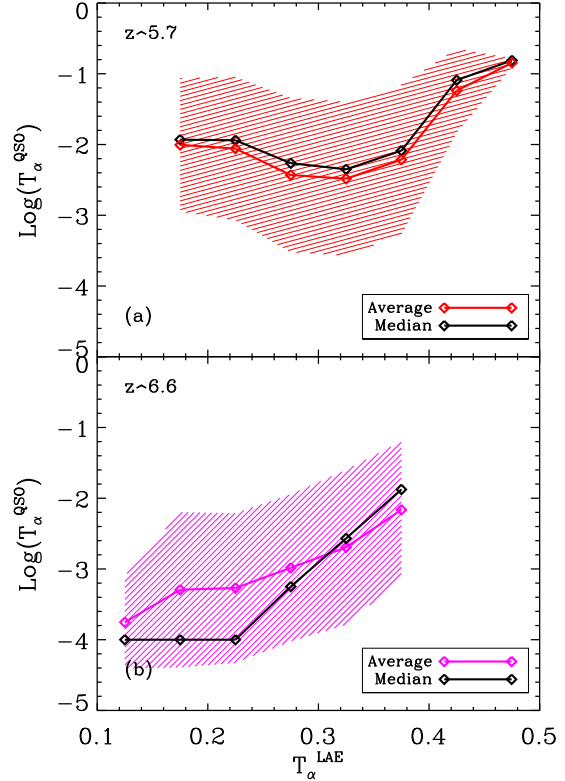}
 \caption{\TQSO (all with impact parameter $0\,<\,d\,<0.5$ cMpc) as a function of \TLAE. \TLAE bins span 0.05 and shaded
 area represent 1$\sigma$ error bars in each \TLAE bin. All \TQSO data below $10^{-4}$ is replace by  $10^{-4}$ which is the actual
 detectable criteria
 } 
 \label{COR1}
\end{figure}

In the previous Sections, we have separately analyzed the properties of \TLAE and \TQSO depending on the relevant physical parameters.
We now ask the question: is there a relationship between the two transmissivities ? To clarify this point, we plot the \TQSO - \TLAE correlation  
in Fig. \ref{COR1}. A general property of the two transmissivities is that \TQSO is sensitive to various physical parameters (e.g.infall velocity,
clustering, neutral fraction) while \TLAE is not. This is because the transmitted photons from QSOs are all in the blue side of the Ly$\alpha$ line,
hence they experience strong scattering in the Ly$\alpha$ profile core as they are redshifted by cosmic expansion. On the other hand, almost
all transmitted photons from LAE are already in the red side of the line having small cross section to \HI atoms when they escape from the galaxy.
Only photons near to the core of the Ly$\alpha$ line are affected by infall or \HI fraction. 

At $z= 5.7$, we do not find a strong correlation between the two transmissivities. Most of the \TLAE data points loiter
in 0.2-0.3, whereas \TQSO shows a large scatter in  $10^{-6}-1$. \TLAE is weakly anti-correlated with 
$v_{inf}$ and this erases the possible correlation with \TQSO, which instead is strongly correlated with the infall velocity. 
However, at $z = 6.6$ there is a clear positive  correlation between the two transmissivities, although the dispersion around it is relatively large.
as shown in panel (b) of Fig. \ref{COR1}.
Almost \TQSO data  below  \TLAE$\le 0.09$ drop under $10^{-4}$ which is the detection criteria, but they increase very sharply from $10^{-4}$ to
$0.1$ between \TLAE$\approx0.2$ and \TLAE$\approx0.4$. 
LAEs with low \TLAE are embedded in the almost neutral IGM patches, and the photons from background QSO passing
in their vicinity have a very high probability to be scattered out of the LOS. 

\section{Summary and discussion}

The evolution of the luminosity function (LF) of high-$z$ galaxies and, in particular, of LAEs is one of the most promising tools
to study the interplay between early galaxy formation and cosmic reionization. However, before solid conclusions can
be drawn from such a method, a number of delicate aspects must be worked out quantitatively.   One of the most challenging ones
is the presence of a degeneracy between Ly$\alpha$ line damping by dust in the source and/or by intergalactic \HI. As 
\citet{Daya11} have pointed out, a wide range of \HI fractions is compatible with the observed Ly$\alpha$ luminosity function
since the effects of a largely neutral IGM are compensated by a higher escape fraction of Ly$\alpha$ photons from 
galaxies containing only small amounts of dust. With the aim of breaking this dust/\HI degeneracy we have proposed
a possible solution which relies on an independent determination of \xhi exploiting the presence of a background 
luminous source (typically a QSO but the use of Gamma Ray Bursts can be also conceived) whose LOS pierces through and samples the
target LAE surrounding matter.  The theoretical basis of the method, which we have applied at two different redshifts 
$z = 5.7$ and 6.6 where large LAE and QSO samples are available, can be summarized as follows. 

First we normalize the UV ionizing flux and the \HI field by using  results from QSO absorption line experiments.
We produce 3000 synthetic QSO spectra at redshift $z=5.7$ with randomly selected 
position and direction in a $(100h^{-1}\text{Mpc})^3$ simulation volume. Next we obtain the transmissivity for each LOS, using the 
same spectral resolution and wavelength range as for a sample of 17 observed QSOs \citep{Song04,Fan06}.
We find that the transmissivity distribution changes very sensitively with the mean UV photoionization rate; 
the best fit to the data is obtained for a volume-averaged $\langle x_{HI} \rangle =2.3\times 10^{-5}$.

Before running Ly$\alpha$ radiative transfer for all identified LAEs in the simulation, we have presented the individual case 
of a prototypical LAE to illustrate some relevant physical features behind the Ly$\alpha$ line transfer. The observed surface 
brightness  ($SB \propto r^{-3}$) is the sum of the central point source and a scattering halo extending up to 150 kpc from the galaxy center. Such halo
is just below current detection threshold and at reach of future experiments. Our results also indicate that Ly$\alpha$ photons 
in the red wing of the line are largely transmitted at this redshift. We also find that infalling gas is the dominant process causing a 
reduction of transmissivity, whereas outflows seem to be efficiently quenched by infall ram pressure, thus confining their 
effects in a relatively small region around the galaxy ($< 50 $ kpc). 

For each resolved LAE in the simulation, we select a small box of $(10h^{-1}\text{Mpc})^3$ around it and derive
the optical depth $\tau$ along the three LOS parallel to the three viewing axes. At redshift $5.7$, we find that 
\TLAE $\approx 0.25$, almost independent of the halo mass. This constancy arises from the conspiracy of two different
physical effects: (i) the intrinsic Ly$\alpha$ line width and (ii) the infall peculiar velocity. The  Ly$\alpha$ width depends 
on the galaxy rotation velocity, which increases with halo mass; the broader is the line, the larger is the fraction of red wing
photons that can be transmitted, thus increasing \TLAE.  The infall velocity also increases with halo mass, but it acts to
suppress \TLAE, as it blue-shifts the line in the rest frame of the infalling \HI atoms. Evidently, the two effects balance each
other almost perfectly, yielding a roughly constant \TLAE. At higher redshift, $z=6.6$, where $\langle x_{HI} \rangle =0.25$ 
the transmissivity is instead largely set by the local \HI abundance and by the ability of the galaxy to carve a sufficiently large 
\HII region around itself. As the SFR and the photoionization rate increases toward large masses, \TLAE consequently 
increases with halo mass from 0.15 to 0.3. 

As a next step, we cast thousands of LOS originating for background QSOs passing through foreground LAEs at different impact 
parameters. Differently from the case of LAE, photons from the quasars are in the blue side of the Ly$\alpha$ line and are scattered out 
of the LOS by \HI in the vicinity of the LAE. Again, we have emphasized the importance of gas infall motions which increase the 
quasar transmissivity (\TQSO). At smaller impact parameters, $d < 1 $ cMpc, a positive correlation between \TQSO and $M_h$ is
found at $z=5.7$, which tends to become less pronounced (i.e. flatter) at larger distances. Quantitatively, a roughly 10$\times$ 
increase (from $5\times 10^{-3}$ to $6\times 10^{-2}$) of \TQSO is observed in the range $\log M_h = (10.4-10.6)$. The correlation is even 
stronger at $z=6.6$.

At $z= 5.7$, we do find a relatively weak correlation between\TLAE and \TQSO.  This is because most of the \TLAE data points loiter
in 0.2-0.3, whereas \TQSO shows a large scatter in  $10^{-6}-1$. However, at $z = 6.6$ there is a clear positive  correlation between the 
two transmissivities, although the dispersion around it is relatively large. 
The median is below $10^{-4}$ for \TLAE$\le 0.09$, and it increases very sharply from $10^{-4}$ to
$0.1$ between \TLAE$\approx0.2$ and \TLAE$\approx0.4$, showing a very strong sensitivity of \TQSO to \TLAE.
The correlation signal is diluted by the fact that the properties of the environment of many LAEs (i.e the most clustered ones) 
is not completely determined by the radiation and peculiar motions caused by the central object alone. Thus a QSO line of sight through 
such a clustered environment results in contaminating high \TQSO data. Properly selecting relatively isolated LAEs would boost the 
correlation signal.

The proposed method therefore appears to be promising to determine the evolution of the physical state of the IGM with redshift and to allow
a detailed study of reionization. Obviously, its feasibility relies on the availability of a sufficient number of high redshift LAEs and QSOs. 
In this sense, we live in an age full of promises. Currently, about 60 QSOs at 5.7 $<\,z\,<$ 6.5 have been detected in total. Of these, 
\citet{Fan06} discovered 19 QSOs with $z\gsim6$ and \citet{Song04} discovered 5 QSOs at $z\gsim5.7$ in the Sloan Digital Sky Survey\footnote{http://www.sdss.org/}   (SDSS) using moderate resolution spectra, $R\approx$ 5000.  
Additional 19 QSOs are discovered by \citet{Will07,Will09,Will10} in the Canada-France High-$z$ Quasar Survey (CFHQS) with moderate resolution
$R\approx1000-5000$. In the VLA FIRST survey\footnote{http://sundog.stsci.edu},  \citet{Beck06,Beck10} have obtained 17 QSOs with high 
($R\approx40\,000$) and moderate ($R\approx$5000) spectral resolution.  Recently \citet{Mort11} discovered the highest QSO at $z=7.085$.  
New QSOs are still being discovered and identifying QSOs at even higher redshifts is the next challenge.  Ongoing near-IR sky surveys like the UKIDSS  
(\citealt{Lawr07}) will be completed in few years. The Large Area Survey (4000 $\text{deg}^2$) aims at finding $z=7$ quasars. Finally, the ESO VISTA
Telescope surveys \citep{Suth09} will be even more efficient building on a larger camera probably capable of finding $z\gsim 8$ sources.

On the LAE side, a large sample already exists at $z\gsim5.7$. Up to now 54 (45) LAEs are confirmed by spectroscopic observation 
at redshift $z=5.7$ (6.5) in the Subaru Deep Field \citep{Kash11}. The number of confirmed LAEs are 70\% (81\%) of their photometric 
candidates at  $z=5.7$ (6.5). The Subaru/XMM-Newton Deep Survey (SXDS) has also found a large number of LAEs at $z=5.7$. In a 
$1\,\text{deg}^2$ sky \citet{Ouch08} found 401 photometric candidates, 17 of which spectroscopically confirmed. The survey locations of 
SDF ($\alpha=13\text{h}24$, $\delta=27^{\circ}$) and SXDS ($\alpha=2\text{h}18$, $\delta=-5^{\circ}$) are within the survey area of VLA 
FIRST and SDSS which provides a first direct opportunity to test our method. 

The survey field for LAE is about 1 $\rm{deg}^2$ in which hundreds of candidates and tens of confirmed LAE have been found. 
On the other hand QSOs are found over wide area over thousands of  $\rm{deg}^2$ and the probability that we can find them
next to the target LAE is low. A much more efficient strategy to find QSO-LAE pairs would be to perform a deep spectroscopic 
sky survey around a known high-$z$ QSO. The quasar J1335+3533 can be an optimal candidate for that, as it is suitably 
located close to the SDF. We will consider these follow-up ideas  in future work. 
\label{sec_CON}

\appendix
\section{Full radiative transfer vs. exponential models}
\label{fulle}
\begin{figure}
 \includegraphics[width=84mm]{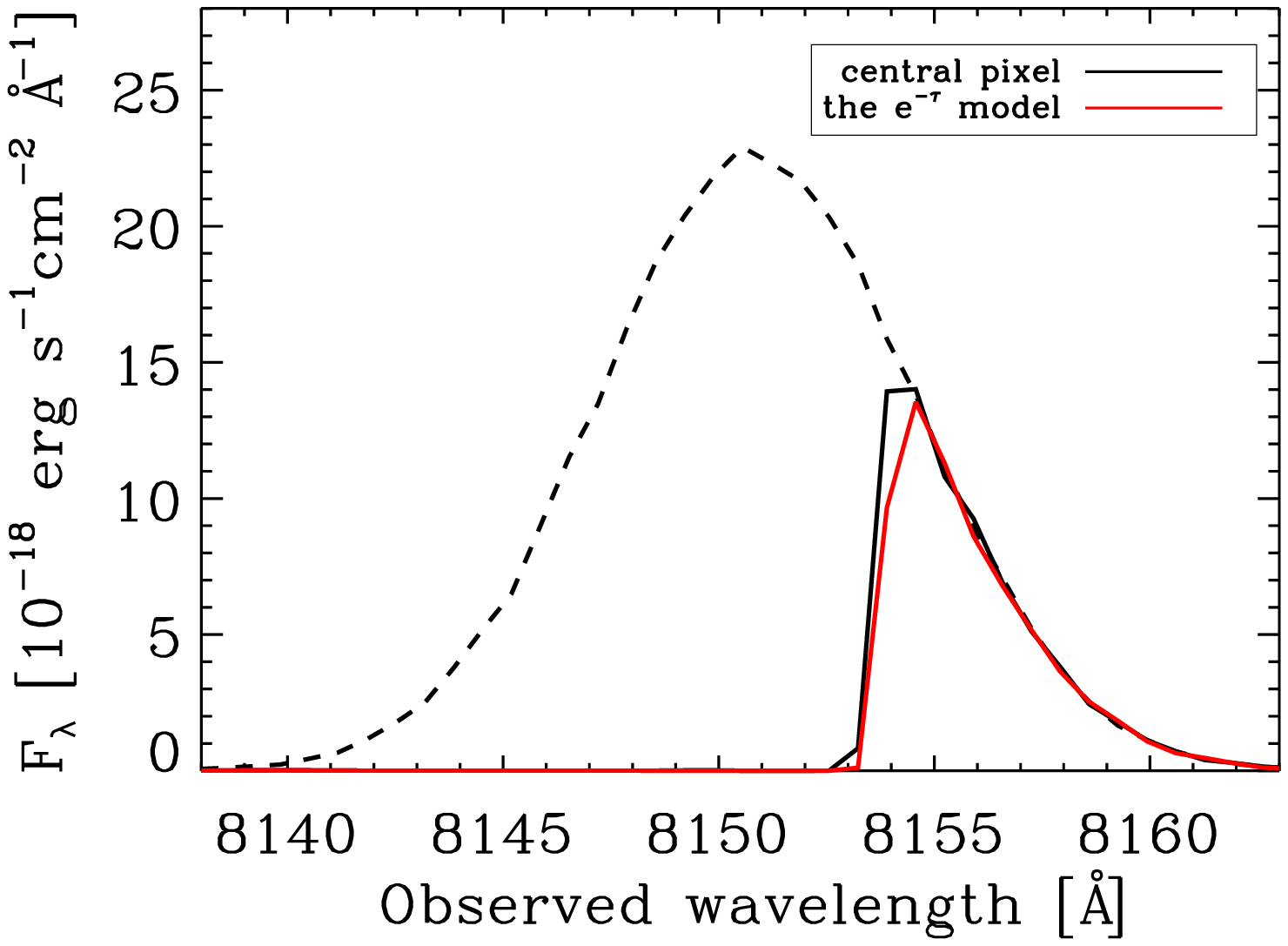}
 \caption{Comparison of the emerging Ly$\alpha$ spectra computed with the full RT (black) vs. exponential (red) models from the central pixel of 
a simulated prototypical LAE at $z=5.7$. The dotted line shows the intrinsic line profile.} 
 \label{compare}
\end{figure}
\citet{Zhen10} have studied LAEs using full Ly$\alpha$ radiative transfer in cosmological reionization simulations. 
They compare their results with previous works (e.g. \citealt{Mcqu07b,Ilie08c, Daya11}), where the transfer of Ly$\alpha$ is 
treated in an approximate manner: in such works, in fact, the intrinsic line profile is damped by $\text{e}^{-\tau_{\nu}}$, 
where $\tau_{\nu}$ is the scattering optical depth at frequency $\nu$ along the LOS. \citet{Zhen10} emphasize that the  
$\text{e}^{-\tau_{\nu}}$ model does not provide an accurate description of the observed Ly$\alpha$ spectrum, because it does not account for 
frequency (no frequency change occurs for any Ly$\alpha$ photon) and space (no surface brightness information) diffusion. 
To investigate further this point, in addition to the full radiative transfer model discussed so far, we have reconstructed the LAE spectra 
also with the exponential model. We can then evaluate the differences between the two. 

The black curve in Fig.\ref{compare} shows the spectrum from the central pixel of the prototypical LAE at $z=5.7$ presented in
Sec. \ref{proto}. The apparent shift of the peak with respect to the intrinsic profile is about 5 \AA.
This is consistent with the result of \citet{Zhen10} (see their Fig.6) for the same halo mass, $\log(M_h) > 10.7$ $h^{-1}M_{\odot}$.
The red curve in the same Figure has been obtained from the $\text{e}^{-\tau_{\nu}}$ model. 

We find that at all frequencies such model underestimates the flux, but different from \citet{Zhen10}, we find the difference between 
the two methods on \TLAE to be $< 1$\%. Most of the scatterings occurring in the central pixel shift the photon frequency in the range 
within $8148\,\text{\AA} < \lambda< 8153\,\text{\AA}$, corresponding to the line core and thus they are efficiently damped. 
Average over all the LAEs identified in the simulation, the transmissivity computed with full radiative transfer is larger by about 
$5\%$; some halos (0.5\% of total) are embedded in a dense HI region, where $n_{\rm{HI}}\ge10^{-4}\rm{cm}^{-3}$. 
Full radiative transfer of these halos gives \TLAE$\approx 0.3-0.4$, while the \emodel model suppresses the line almost completely.

We have analyzed the possible reasons for the discrepancy between \citet{Zhen10} and our findings. We came to the conclusion that 
it originates from the combination of two factors: the different line width and IGM properties. 
\citet{Zhen10} used line widths 
\begin{equation}
\frac{\Delta \nu_D}{\nu_0} =10^{-4} [M_h/(10^{10}h^{-1}M_{\odot})]^{1/3}.
\end{equation}
This is about 3 times smaller than what adopted here.
With this line width, almost all the photons are initially within the core of the line profile and the flux of the  $\text{e}^{-\tau_{\nu}}$ model is
greatly suppressed; as a result only a
very small amount of photons can escape without any scattering and the main contribution to the observed flux is coming from scattered photons.
In our case, on the other hand,  most photons with $\lambda>8155$ \AA \, are transmitted without any scattering and 
only a  very small contribution is coming from scattered photons. Even when we enlarged the  integrating flux area to $r\leq50$ kpc the difference 
with respect to the \emodel model remains $< 5$\%. The very small contribution from scattered photons is related to a different HI density field. 
As discussed above we have carefully calibrated the IGM transmissivity  through an iterative optimization with QSO absorption spectra. This procedure gives 
on average a 4 times lower $\langle n_{\text{HI}} \rangle$ than adopted by \citet{Zhen10}, which results in a reduced spatial/frequency diffusion. 

Finally, the neutral density at the very center of halo can be underestimated due to the lack of mass resolution. The cell size of Ly$\alpha$ radiative 
transfer is 1.6 times smaller than \citet{Zhen10} but the mass resolution is 1700 times larger than theirs.  Hence, we have less particles per halo, 
potentially leading to underestimating the  recombination rate and $n_{\rm{HI}}$.  \TLAE is determined by the amount of transmitted photons rather than
scattered photons.
However, we verify that even with 10 times higher  \HI density in 
the central cell, the photons with $\lambda >8155$ \AA\  are transmitted without scattering since they are far from the core of the line profile.  
The fraction of photons transmitted without scattering events is rather sensitive to the peculiar velocity of the 
infalling gas. As we shown in Fig.\ref{velo}, however, the infall profile is well captured by remapping SPH particles on a finer grid.

\section*{Acknowledgments}
We acknowledge useful discussions with B. Ciardi, A. Mesinger, P. Laursen and other DAVID members.

\end{document}